\documentclass[twoside,12pt]{article} % had been art12 
\font\runningheadfont=cmcsc10 

\def\refp#1{(\ref{#1})}

\def\lrp#1{\left( #1\right)} 
 \newif\ifMarginNotes \MarginNotestrue 
\def\mrgn#1{\ifMarginNotes\setbox0=\vtop{\hsize 6.75pc 
   {\noindent\relax #1\par}}\leavevmode 
   \vadjust{\dimen0=\dp0 \dimen1=\ht0\advance\dimen1 by .5ex 
 \advance\dimen0 by -.5ex 
  \kern-\dimen1\hbox{\kern\hsize\kern.5pc$\leftarrow$ 
  \box0}\kern-\dimen0}\fi} 
\catcode`\@=11 
\font\twelvemsb=msbm10 scaled 1200 
\font\tenmsb=msbm10 
\font\ninemsb=msbm7 scaled 1200%msbm9 
\newfam\msbfam 
\textfont\msbfam=\twelvemsb  \scriptfont\msbfam=\tenmsb 
  \scriptscriptfont\msbfam=\ninemsb 
\def\msb@{\hexnumber@\msbfam} 
\def\Bbb{\relax\ifmmode\let\next\Bbb@\else 
 \def\next{\errmessage{Use \string\Bbb\space only in math 
mode}}\fi\next} 
\def\Bbb@#1{{\Bbb@@{#1}}} 
\def\Bbb@@#1{\fam\msbfam#1} 
 
\font\twelveeufm=eufm10 scaled 1200 
\font\teneufm=eufm10 
 
\font\seveneufm=eufm7 
 
\newfam\eufmfam 
\textfont\eufmfam=\twelveeufm 
\scriptfont\eufmfam=\teneufm 
\scriptscriptfont\eufmfam=\seveneufm 
\def\frak{\relax\ifmmode\let\next\frak@\else 
 \def\next{\errmessage{Use \string\frak\space only in math 
mode}}\fi\next} 
\def\frak@#1{{\frak@@{#1}}} 
\def\frak@@#1{\fam\eufmfam#1} 
 
\catcode`\@=12

\def\ritem#1{\item[{\rm #1}]} 
\def\sing{\Upsilon_{{\rm sing}}} 
\def\mbni#1{\vskip18truept\noindent{\bf #1}}

\def\Vrt#1{\Vert #1\Vert} 
\def\la{\lambda} 
 
\def\part{\partial} 
\def\CS{{\cal S}} 
\def\cs{{\cal S}} 
\def\K{{\cal K}}

\def\Tr{{\rm Tr} } 
\def\Trh#1{{{\rm Tr}}_{\fh} \l( #1 \r) } 
\def\llangle{\left\langle} 
 
\def\rrangle{\right\rangle} 
\def\lra#1{\llangle #1\rrangle} 
\def\lrag#1{{\llangle #1\rrangle}_{\gamma} }

\def\hensp#1{\enspace \hbox{#1}\enspace} 
 
\def\la{\lambda} 
\def\th{\theta} 
\def\uth{U(\th)}

\def\C{{\Bbb C}} 
\def\CC{{\cal C}} 
\def\D{{\cal D}} 
\def\R{{\Bbb R}} 
\def\hattorus{{\hat{\Bbb T}^s}}
\def\torus{{{\Bbb T}^s}}
 
\def\CH{{\cal H}} 
\def\ep{\epsilon}

\def\fz{{\frak Z}} 
\def\fc{{\frak c}}
\def\FC{{\frak C}}

\def\fh{{\frak H}} 
\def\Z{\Bbb Z} 
\def\T{\Bbb T}

\def\v#1#2#3{\varphi_{#1}(#2,#3)}
\def\ov#1#2#3{\overline{\varphi_{#1}}(#2,#3)}
\def\opart{\overline{\partial}}
\def\og{{\overline{\gamma}}}
\def\oz{\overline{z}}
\def\clips{,\ldots} 
\def\be{\begin{equation}} 
\def\ee#1{\label{#1}\end{equation}} 
\def\beq{\begin{eqnarray}} 
\def\nn{\nonumber} 
 
\def\eeq{\end{eqnarray}} 
\def\l{\left} 
\def\r{\right} 
 
\def\w{\Omega} 
\def\wj{\w_j} 
\def\wp{{\omega_\theta}}  
\def\wpj{{\omega_{j, \th}}}  
\def\ow{\overline{\wp}}
\def\wg{{\omega_\gamma}}  
\def\wgj{{\omega_{j, \th}}}  
\def\st{{\cs_{\th }}}

\def\spt{{\cs'_{\th} }}

\def\zj{z_j}

\def\dmug{{d\mu_\gamma}}
\def\dmuge{{d\mu_{\gamma,\ep}}}
\def\dmugp{{d\mu_{\gamma'}}}
\def\cmug{{C_\gamma(t,s)}}
\def\cmgu{{C_\gamma(s,t)}}

\def\lraug#1{{\lra { #1 }_\gamma}}
\def\lrauge#1{{\lra { #1 }_{\gamma,\ep}}}
\def\gep{{\gamma,\ep}}

\def\sumj{\sum_{j=1}^n}
\def\sumk{{\sum_{k\in\hat\torus}}}
\def\prodj{\prod_{j=1}^n}
\def\lrdq#1 {{\vdots {#1} \vdots_\gamma }}
\def\lrd#1 {{ \vdots {#1} \vdots }}
\def\tt{{\tau,\th }}
\def\utth{U(\tt)}

\def\Pj{\Phi_{\tt,j}}
\def\Ph{\Phi_{\tt}}
\def\Pjxt{\Phi_{\tt,j}(x,t)}
\def\Pxt{\Phi_{\tt}(x,t)}
\def\P#1#2#3{\Phi_{\tt, #1}(#2,#3)}
\def\oP#1#2#3{\overline{\Phi_{\tt, #1}(#2,#3)}}
\def\green{G_{\th}}

\def\vol{{\rm Vol}}
\def\slim#1{\mathop {\rm st.\,lim}_{#1}}
\def\hm{H^{(m)}}
\def\hmp{H^{(m^\prime )}}
\def\ehm#1{e^{- {#1} \hm}}
\def\ehmp#1{e^{- {#1} \hmp}}
\def\intbeta{\int_{0}^{\beta} }
\def\vac{\w_{\rm vacuum}}

\title{Twist Positivity\thanks{Work supported 
in part by the Department of Energy under Grant DE-FG02-94ER-25228 and 
by the National Science Foundation under Grant DMS-94-24344.}} 
\author{Arthur Jaffe\\ 
Harvard University\\ 
Cambridge, MA 02138, USA} 
\date{August 1998} 
\begin{document} 
\hsize=7truein 
\hoffset=-.75truein  
\maketitle 
\thispagestyle{empty} 
 
\begin{abstract} 
We study a heat kernel $e^{-\beta H}$ defined by a self-adjoint
Hamiltonian $H$ acting on a Hilbert space $\fh$, 
and a unitary representation $U(g)$ of a symmetry group $G$ 
of $H$, normalized so that the ground vector of $H$ is invariant under
$U(g)$. The triple $\{H, U(g),\fh\}$ defines a twisted partition function 
$\fz_g$ and a twisted Gibbs expectation $\lra{\ \cdot\ }_g$,
$$
\fz_{g} = \Tr_\fh\l( U(g^{-1})e^{-\beta H}\r)\;,
\quad \hensp{and}\quad
\lra{\ \cdot\  }_{g} = 
\frac{\Tr_\fh\l(U(g^{-1})\ \cdot\ e^{-\beta H}\r)}
{\Tr_\fh\l(U(g^{-1})e^{-\beta H}\r)}\;.
$$ 
We say that $\{H, U(g),\fh\}$ is {\it twist positive} if $\fz_g>0$.
We say that $\{H,U(g),\fh\}$ has a Feynman-Kac representation with 
a twist $U(g)$, if one can construct a 
function space and a probability measure $d\mu_{g}$ on 
that space yielding (in the usual sense on products of coordinates)
$$ 
\lra{\ \cdot\  }_{g } = \int \ \cdot \ d\mu_{g }\;.
$$
Bosonic quantum mechanics provides a class of specific 
examples that we discuss.  We also consider
a complex bosonic quantum field $\varphi(x)$ defined on a 
spatial $s$-torus $\torus$ and with a translation-invariant Hamiltonian. 
This system has an $(s+1)$-parameter abelian twist group $\torus\times\R$,
that is twist positive and that has a Feynman-Kac representation. Given 
$\tau\in\torus$ and $\th\in\R$, the corresponding paths are
random fields $\Phi(x,t)$ that satisfy the {\it twist relation}
$$
\Phi(x,t+\beta) = e^{i\w\th} \Phi(x - \tau,t) \;.
$$
We also utilize the twist symmetry to understand 
some properties of ``zero-mass'' limits, when the twist
$\tt$ lies in the complement of a set $\sing$ of singular twists.
\end{abstract}

\newpage 
\tableofcontents
\newpage
\section{Introduction} 
We study a bosonic quantum mechanical system defined 
on a Hilbert space $\fh$.  We assume that $H$ is a self-adjoint
Hamiltonian with a unique ground state $\vac$, and that the 
heat kernel $e^{-\beta H}$ is trace class. Let $G$ denote a symmetry
group of $H$, and $U(g)$ a continuous unitary representation of $G$ on 
$\fh$, so $U(g)H=HU(g)$. 
Thus $\vac$ is an eigenvector of $U(g)$; normalize the
phase of $U(g)$ so that 
\be
U(g)\vac = \vac\;.
\ee{PhaseNormalization}
This triple defines a twisted partition function $\fz_g$
and a twisted Gibbs expectation $\lra{\ \cdot\ }_g$,
\be
\fz_{g} = \Tr_\fh\l( U(g^{-1})e^{-\beta H}\r)\;,
\quad \hensp{and}\quad
\lra{\ \cdot\  }_{g} = 
\frac{\Tr_\fh\l(U(g^{-1})\ \cdot\ e^{-\beta H}\r)}
{\Tr_\fh\l(U(g^{-1})e^{-\beta H}\r)}\;.
\ee{TwistedGibbsFunctional}
Define $\{H, U(g),\fh\}$ to be {\it twist positive} if for all $g\in G$ 
and for all $\beta>0$,
\be
\fz_{g}  > 0\;.
\ee{TwistPositive}
Clearly $\fz_{\rm Id}>0$. Note that 
$\lim_{\beta\to\infty}\l(\fz_g/\fz_{\rm Id}\r) = \lra{\vac,U(g)\vac}$, 
showing \refp{PhaseNormalization} to be necessary.  

A Feynman-Kac measure is a countably-additive, Borel probability 
measure $d\mu_{g }$. 
We say that $\{H,U(g),\fh\}$ has a Feynman-Kac representation with 
a twist $U(g)$, if one can construct a 
function space and a Feynman-Kac measure $d\mu_{g}$ on 
that space yielding 
\be
\lra{\ \cdot\  }_{g } = \int \ \cdot \ d\mu_{g }\;
\ee{BasicFK}
in the sense that Gibbs expectations of time-ordered products
of coordinates equal the integral of the same product of paths.  
We give examples in Propositions III.3 and V.\refp{NonGaussianFK} for quantum 
mechanics and in Propositions VI.6 and VI.8 for quantum fields.

Once we establish the existence of a measure,
we can use classical inequalities and harmonic analysis to 
study particular integrals, and therefore
to gain quantitative insights concerning $\fz_g$ or $\lra{\ \cdot\ }_{g}$. 
Hence if we have a measure, 
then we may benefit from these basic tools of mathematical 
physics.  Such inequalities are especially useful 
in understanding non-Gaussian expectations
that cannot be evaluated in closed form.\footnote
%%%
%%%
{In problems with fermions, it is common to define a functional 
on a Grassmann algebra, either using Berezin integration or using 
expectations on a Fock space. In specific examples of interest, 
these methods often do not yield positive functionals. Then the resulting 
theory is restricted to the 
Gaussian case, where the functionals yield Pfaffians and determinants. 
In the case of boson-fermion systems the purely bosonic part of the
system may have a Feynman-Kac measure. This measure, multiplied by
a Pfaffian or a determinant, can be studied using the positivity 
properties of the bosonic measure. This method to study boson-fermion 
systems is common in constructive quantum field theory. 
Of course one can combine the measures we discuss in this 
paper with Pfaffians or determinants arising from fermionic degrees of
freedom.}
%%%
%%%

Twist positivity is necessary for $\fz_{g}$ to have a Feynman-Kac 
representation, but it is neither a necessary nor 
sufficient for $\lra{\ \cdot\  }_{g } $ to have a representation.\footnote
{A corresponding necessary condition for $\lra{\ \cdot\  }_{g } $  to have a 
Feynman-Kac representation exists:  the pair correlation operator $C_{g}$, 
defined in \refp{PCop} and in \refp{PCOperator}, should be a linear 
transformation with positive spectrum. 
In the case that $\lra{\ \cdot\  }_{g } $ is a Gaussian functional, 
this condition is also sufficient.}
The fact remains that in the examples that we study, 
twist positivity holds.  Twist positivity 
also appears in our proof of the existence of 
the Feynman-Kac measure for $\lra{\ \cdot\  }_{g } $.  Thus twist 
positivity appears to be an excellent condition on which to focus, and
we make it our title.

Our results apply to expectations that occur various 
branches of physics and of mathematics, including quantum mechanics, 
quantum field theory, and related problems in probability theory. 
We illustrate them with examples, starting in the simplest cases, and
bosonic quantum mechanics provides specific ones. 
Then we discuss other examples from 
quantum field theory.\footnote
{Surprisingly, we have not found these observations in the literature, 
though some of them are ``folk theorems'' in physics,
where their mathematical status is not clarified. }
Consider a quantum theory with a complex coordinate $z$ acting 
as a multiplication operator on the Hilbert space $\fh= L^2(\C)$.
Given a positive constant $\w>0$, define the elementary $U(1)$ twist group 
with period $2\pi/\w$ by 
\be
z\to e^{i\w\th}z\;.
\ee{CoordinateTwist}
Assume that the unitary operators $U(\th)$ acting on $\fh$ implement this 
twist, so $\uth z = e^{i\w\th}z\uth$. Furthermore, assume that the
Hamiltonian $H$ of the system is 
twist-invariant, and has a trace-class heat kernel.  We consider the
Gibbs functional $\lra{\ \cdot\  }_{\th }$ with $\th$ replacing $g$.

In the quantum mechanics case, the measure $d\mu_{\th }$ 
is concentrated on continuous paths. 
H{\o}egh-Krohn discovered that for $\th=0$, notably when 
$\lra{\ \cdot \ }_0$ is a Gibbs state, that a 
Feynman-Kac representation arises from  
paths $\omega_0(t)$ that are periodic in time \cite{HoeghKrohn},
\be
\omega_0(t+\beta) = \omega_0(t) \;.
\ee{periodicpaths}
Our present results generalize this picture. In the simplest case of 
quantum theory with the twist \refp{CoordinateTwist}, the twisted 
Feynman-Kac representation \refp{BasicFK} arises from paths that are 
twisted periodic, namely they satisfy the twist relation 
\be
\wp(t+\beta) =  e^{ i\w\th} \wp(t) \;.
\ee{twistedperiodicpaths}
In the case of a vector-valued coordinate $z\in\C^n$,  
we may use a different rate of twisting in each coordinate 
direction.  Thus we replace \refp{CoordinateTwist} by the relation
\be
z_j \to \uth z_j \uth^\ast = e^{ i\wj \th} z_j\;,
\ee{VectorCoordinateTwist}
with individual periods of twisting $2\pi/\wj>0$, for each
$1\le j\le n$. We call the  $\w=\{\wj\}$ {\it weights}.
In place of \refp{twistedperiodicpaths}, our 
vector-valued paths satisfy the twist relation 
\be
\wpj(t+\beta) = e^{ i\wj\th} \wpj(t) \;, \quad\hensp{for}\quad
1 \le j\le n\;.
\ee{VectorTwistedPeriodicPaths}
The existence of a minimal, strictly-positive constant $h$ such 
$h\wj\in\Z$ for all $1\le j\le n$, is equivalent to 
the ratios of weights being rational.  In that case, 
we say that the weights $\{\wj\}$ are rationally commensurate. The 
group $U(\th)$ then has a period $2\pi h$.

In \S II--IV we study a
twist-invariant oscillator with Hamiltonian $H=H_0$ and frequency
$m>0$. We work out this example in great detail, because our quantum
mechanics results for other potentials rely on a detailed understanding 
of the oscillator.
We give an elementary argument that the twisted Gibbs functional 
for the oscillator is Gaussian, and we give the Feynman-Kac measure
$d\mu_{m,\beta,\th}$. This measure is Gaussian and it is twist-invariant.
The covariance of the measure is 
the pair correlation function of the oscillator. 
In Theorem III.1 we identify the covariance 
as a the resolvent (i.e. Green's function) of a twisted Laplacian.  
Also we show that this pair correlation function
extends as a function of the time difference $t-s$ into the complex plane, 
to be complex periodic. The imaginary part of the period determines the 
twist angle $\th$. 

Once we have derived properties of the twist-invariant oscillator,
we study perturbed Hamiltonians $H$ of the form 
\be
H = H_0 + V\;.
\ee{TwistInvariantHamiltonian}
Here $V$ is a suitable real, twist-invariant function that is 
bounded from below. 
We detail our assumptions on $V$ in Definition V.1. 
In order to make this situation more concrete, let us illustrate some 
sorts of acceptable potential functions $V$.  
\begin{itemize}
\item[{\bf  V-i}] The absolute 
square $V(z, \oz) = |W(z)|^2$, of a holomorphic, homogeneous 
polynomial $W(z)$ is acceptable.
\ritem{{\bf  V-ii}} A sum of acceptable potentials $V$'s is an acceptable
potential.
\end{itemize}
One acceptable such sum is the square of the gradient of a holomorphic, 
homogeneous polynomial $W$,
\be
V(z, \oz) = \sumj \l|\frac{\part W(z)}{\part z_j}\r|^2\;.
\ee{supersymmetryV}
Such $V$'s arise as the bosonic potential in quantum theory with $N=2$ 
supersymmetry, where $W(z)$ is called the superpotential.  (The polynomial
$W$ may be quasihomogeneous in place of homogeneous, see \S V.)

The twisted Gibbs expectations that arise from the Hamiltonians $H=H_0+V$
are non-Gaussian.  They possess Feynman-Kac representations, and the 
measures defining these representations are non-Gaussian.  The measures 
leading to \refp{BasicFK} have the form
\be
d\mu_{m,\beta,\theta}^{V} 
  = \frac{e^{-\int_0^\beta V(\wp(s), \,\overline{\wp(s)})ds}
       d\mu_{m,\beta,\theta}}
{\int e^{-\int_0^\beta V(\wp(s), \,\overline{\wp(s)})ds}
     d\mu_{m,\beta,\theta}}\;.
\ee{nongaussian}
The normalizing factor for these measures is in fact the ratio of two 
traces, and we define this ratio to be the relative twisted partition 
function of $H = H_0 +V$ relative to $H_0$, 
\be
\fz^V_{m,\beta,\th} 
= \frac{\Tr_\fh \l(\uth^* e^{-\beta (H_0 +V)}\r) }
            {\Tr_\fh \l(\uth^*  e^{-\beta H_0}\r)  }
= \int e^{-\int_0^\beta V(\wp(s),\,\overline{\wp(s)})ds}
     d\mu_{m,\beta,\theta}\;.
\ee{RelativePartitionFunction}
We find that these partition functions are positive for all $\th$,
and for all $\beta>0$, namely
\be
0 < \Tr_\fh \l(\uth^*  e^{-\beta H}\r) \;,\qquad\hensp{and so}\qquad 
0 < \fz^V_{m,\beta,\th}\;.
\ee{PositiveZ}
Furthermore, inspecting these representations shows that if $V$ grows 
at least quadratically in $|z|$ as $|z|\to\infty$, then  
\be
0 < \Tr_\fh\l(\uth^*  e^{-\beta (H_0 + \la^2 V)} \r) 
  \le \l(\frac{M}{\beta(m+\la)} \r)^{2n}\;,\quad\hensp{for}\quad
0\le\la\le1 \;,
\ee{CanonicalZBound}
where M is a constant depending only on $V$.

Functional integration is a central tool for 
quantum field theory, for string theory, and for statistical mechanics. 
The study of zero-mass measures poses special difficulties; 
they occur both in the Hilbert space the Hilbert space formulation of 
field theory and also in the formulation by functional integration. 
We clarify certain aspects of zero-mass limits by using twisted 
expectations in the complex case. 
A paper of Witten \cite{Witten} suggests some properties of 
this sort, the understanding of which led us to this investigation.  
We plan to use this framework in another paper 
that provides quantum mechanics and field theory examples
to illustrate the phenomena in \cite{QHA}.

Define the singular set of twisting angles  
\be 
\sing=\{\th : e^{i\wj\th} = 1\;, \ \hensp{for any}\ 1 \le j\le n\}\;.
\ee{sing}  
If the weights are rationally commensurate, the set $\sing$ contains 
a finite number of points for $\th$ in a period interval.
We prove that for fixed $\beta$ and for fixed $\theta\not\in\sing$ 
the twisted zero-mass measure exists as a weak limit. As $m\to0$,  
\be 
d\mu_{m,\beta,\theta}^V \stackrel{\rm w}{\longrightarrow} 
d\mu_{0,\beta,\theta}^V\;. 
\ee{limit} 
For each $\beta$, $\theta$, the limiting measure 
$d\mu_{0,\beta,\theta }^V$  
is a countably-additive, Borel probability measure.  These results improve 
on the canonical bounds \refp{CanonicalZBound}, 
and in this case the twisted trace 
$\Tr_\fh\l(\uth e^{-\beta (H_0+\la^2V)}\r)$ converges for 
$\th\not\in\sing$ when  $m + \la \to 0$ with $\beta, \th$ fixed.

In addition our method applies to quantum fields $\varphi(x)$, that 
may also have vector components, $\varphi(x)=\{\varphi_j(x)\}$. We take the
spatial variable $x$ in an $s$-torus $\torus$. We take the
spatial periods equal to $\ell_i>0$, where $1\le i\le s$, and let the spatial
volume be $\vol = \prod_{i=1}^s \ell_i$.
Then the Fourier decomposition of the field has the form 
\be
\varphi_j(x) = \frac{1}{\sqrt{\vol}}
 \l(z_j + \sum_{k\neq 0} \hat\varphi_j(k)e^{-ikx}\r)\;, 
\ee{field}
where $k=\{k_i\}$ ranges over the lattice 
$k_i\in\frac{2\pi}{\ell^i}\Z$ dual to $\torus$.  
In the  case that $\varphi(x)$ is a complex field, the 
constant Fourier modes $z_j$ are just the complex coordinates in quantum
theory considered above.  Let $H$ denote a twist-invariant,
translation-invariant Hamiltonian for the field. We assume
that the Hamiltonian operator $H$, the momentum operator $P$, 
and the symmetry operator $\uth=e^{i\th J}$ mutually commute. 
We also assume that $H$ has a unique ground state
$\vac$, so $\vac$ is an eigenstate of $\uth$ and of the momentum.
We take  
\be
\uth\vac=\vac\;,
\quad\hensp{and}\quad
P\vac =0\;.
\ee{VacuumTwistRenormalization}
We consider $\uth$ implementing a twist similar to before, 
\be
\varphi_j(x) \to \uth \varphi_j(x) \uth^\ast = e^{i\wj\th} \varphi_j(x)\;.
\ee{QuantumFieldTwist}
With these assumptions, we
can interpret the translation group as an additional twist, yielding 
the $(s+1)$-parameter twist group 
\be
U(\tt) = e^{i\tau P +i\th J} = e^{i\tau P}\uth = \uth e^{i\tau P}\;.
\ee{sTwist}
Here $U(\tau,0)=e^{i\tau P}$, with $\tau\in\torus$. In terms of  
components, $\tau P = \sum_{i=1}^s \tau_i P_i$.  Then
\be
\utth\varphi_j(x)\utth^* = e^{i\wj\th} \varphi_j(x - \tau)\;.
\ee{GeneralQuantumFieldTwist}
Twist positivity is the statement 
\be
\fz_{m,\beta,\tt} = \Tr_\fh \l( U(\tt)^* e^{-\beta H} \r)
=\Trh{\uth^*  e^{-i\tau P - \beta H}} > 0\;.
\ee{fieldTwistPositivity}
The limit $\beta\to\infty$ shows \refp{VacuumTwistRenormalization} is
necessary. 
Define the twisted Gibbs functional 
\be 
\lra{\ \cdot \ }_{m,\beta,\tt}  
= \frac{\Tr_{\fh}(U(\th)^* \ \cdot\  e^{-i\tau P - \beta H})}
{\Tr_{\fh}(U(\th)^* e^{-i\tau P - \beta H})}\;. 
\ee{fieldGibbs} 
We first establish this condition for the massive free field, with $H=H_0$,
that twisted positivity holds.  
This allows us to prove that the massive free-field pair correlation 
operator has strictly positive spectrum.  Furthermore, we show that the 
twisted free-field Gibbs functional is Gaussian.  
These two properties lead to a Feynman-Kac representation for the 
corresponding free-field functional given by a measure $d\mu_{m,\beta,\tt}$.

The probability interpretation for the quantum field case arises from 
paths called random fields, and defined on 
a space-time $\CC = \torus \times [0,\beta]$.
Since the random fields satisfy a twist relation
depending on both $\tau$ and on $\th$, we denote them 
$\Pxt$ with components $\Pjxt$, for $1\le j\le n$. 
%We also refer to this 
%boundary condition as a {\it twist relation}.
After averaging with a smooth function of the coordinate $x$, the 
random fields give paths that are continuous functions of the time. 
Furthermore, the translation group acts continuously on $C^\infty$ 
functions on $\CC$, also in the Fr\^echet topology, so translations also act 
continuously on random fields.  We find in \S VI that the appropriate twist 
relation for random fields is 
\be
\Pj(x, t+\beta) = e^{i\wj\th} \Pj(x - \tau,t) \;.
\ee{RFTwistRelation}
There is a Gaussian Feynman-Kac representation for the massive free-field 
Hamiltonian $H_0$ with the above twist.  We denote the Gaussian measure by 
$d\mu_{m,\beta,\tt}(\Ph(\cdot))$.  We show that the covariance of this 
measure $C_\tt$ equals the resolvent of a twisted Laplace operator 
$\Delta_\tt$, 
\be
C_\tt = (-\Delta_\tt + m^2)^{-1}\;.
\ee{CovarianceLaplace}
We also construct non-Gaussian measures 
$d\mu_{m,\beta,\tt}^{V}(\Ph(\cdot))$ for 
interacting fields with certain Hamiltonians $H = H_0 + V$. These measures
provide 
Feynman-Kac representations for twisted Gibbs functionals \refp{fieldGibbs}.
We regularize these theories to avoid discussing renormalization at this 
time. The measures have the form 
\be
d\mu_{m,\beta,\tt}^{V}(\Ph(\cdot)) 
  = \frac{e^{-\int_\CC V(\Phi_{\tt,\chi}(y,s), 
    \overline{\Phi_{\tt,\chi}(y,s)}) ds dy }
d\mu_{m,\beta,\tt}}
{\int e^{-\int_\CC V(\Phi_{\tt,\chi}(y,s), 
    \overline{\Phi_{\tt,\chi}(y,s)}) ds dy }
     d\mu_{m,\beta,\tt}}\;.
\ee{nongaussianField}
Here $\chi$ denotes that the random field $\Phi_{\tt,\chi}$ has a 
regularization, as is the potential in $H=H_0+V$.
In \S VI, we establish a result concerning the $m\to0$ limit for non-sinular
twisting for the complex quantum field.

In \S VI we also study real fields $\varphi(x)$. 
These fields still have the 
$s$-parameter group arising from translations that is twist positive.  
The corresponding random fields depend on the parameters $\tau$, but not on 
$\th$ and we denote them $\Phi_\tau(x,t)$.  Given $\tau\in\torus$, 
these random fields satisfy
\be
\Phi_\tau(x, t+\beta) = \Phi_\tau(x - \tau,t)\;.
\ee{RealTwist}
They also lead to a twist-positive partition function and a Feynman-Kac
representation.

\goodbreak
\section{The Twist-Invariant Oscillator} 
\setcounter{equation}{0}
We define a twist-invariant harmonic oscillator as a harmonic oscillator 
with a complex coordinate and a rotationally symmetric potential, equal to 
$m^2|z|^2$.  
We now study the complex oscillator in detail. 

\subsection{Canonical Field Coordinates}
We assume that the frequency (mass) of the oscillator is strictly positive,
$m>0$, use the normalization suggested by field theory, in which the 
coordinate $z$ is proportional to $m^{-1/2}$, and linear in dimension-less 
creation and annihilation operators.

Assume $z$ take values in $\C$, and let $\fh = L^2(\C,dz)$ denote
the Hilbert space of $L^2$ functions with respect to
Lebesque measure on $\C$.  Let $\part = {\part\over \part z}$ and 
let $\opart = {\part\over\part \oz}$, so $\opart=-\part^\ast$.  
We introduce independent annihilation operators $a_+$ and $a_-$, and write 
\be 
a_+  =  {1\over \sqrt 2} (\sqrt{m} \oz+{1\over \sqrt m} \part)\;,
\hensp{and} a_- = {1\over \sqrt 2} (\sqrt{m} z+{1\over \sqrt m} \opart)\; .
\ee{adef}
Let  $a^\#$ denote  either $a$ or $a^\ast$.  
As a consequence of \ref{adef}, we find the canonical commutation relations 
(CCR) for the operators $a_\pm$ and their adjoints.
\be
[a_\pm,a^\ast_\pm] = I, \hensp{and} [a_\pm,a_\mp^\#]=0 
\ee{ccr}
Also \refp{adef} inverts to yield 
\be
z = {1\over \sqrt{2m}} (a^\ast_+ + a_-)\;, \qquad
\part = \sqrt{m\over 2} (a_+-a^\ast_-) \; .
\ee{coord}
The Hamiltonian $H=H_0$ has the form
\beq
H_0 &=& \part^\ast \part +m^2 |z|^2-mI=-\part\opart +m^2|z|^2 - mI \nn\\
&=& m \lrp{ a^\ast_+ a_+ +a^\ast_- a_-}\; , \label{ham}
\eeq
where we subtract the constant $m$ to ensure that the minimum eigenvalue 
of $H_0$ equals $0$.  Let
\be
N_\pm = a^\ast_\pm a_\pm\; , 
\ee{number}
denote two commuting number operators.  They have non-negative, integer 
spectrum, and we denote the eigenvalues by $n_+,n_-$  
respectively.  We may also write 
\be
H_0=m(N_++N_-)\; 
\ee{hamnum}

The generator $J$ of the symmetry $U(\theta)=e^{iJ\th}$ 
has a simple form, either in terms of the coordinates, or in 
terms of $N_\pm$, namely
\be
J=\w (z \part - \oz\opart) = \w(N_+-N_-)\; .
\ee{jdef}
Then 
\be
\uth z \uth^* = e^{i\w\th} z\;, \quad \hensp{and} \quad
\uth \partial \uth^* = e^{- i\w\th} \partial\;.
\ee{zsymmetry}
These representations of $H$ and $J$ in terms of $N_\pm$ show that they 
commute as self-adjoint operators. Consider their joint spectrum  $\Sigma$; 
we represent a point in  $\Sigma$ as
\be
h=m(n_++n_-) , \quad n=\w(n_+-n_-)\; . 
\ee{spectrum}
Each point is specified by a unique pair of non-negative integers 
$\{n_+, n_-\}$ that are eigenvalues of $N_\pm$,
and each point in the joint spectrum has multiplicity one.  
Let $|n_+,n_-\rangle$ label an orthonormal basis of simultaneous 
eigenvectors.

Let us introduce the complex constant  $\gamma$, lying in the
interior of the unit disc.  This parameter encodes most of the dependence 
on $m,\beta,\th$ and $\w$.  Let
\be
\gamma=e^{-m\beta+i\w\th}\;. 
\ee{q}
Then we denote our twisted partition function by $\fz_{\gamma}$ and our 
twisted expectation by $\lra{\,\cdot\,}_\gamma$, where 
\be
\fz_\gamma = \Tr_{\fh} \l(U(\theta)^* e^{-\beta H_0}\r) \;,
\quad\hensp{and}\quad
\lra{\,\cdot\,}_{\gamma} 
= \frac{\Tr_{\fh}(U(\theta)^* \, \cdot\, e^{-\beta H})}
       {\Tr_{\fh}(U(\theta)^* e^{-\beta H})}\;.
\ee{qfunctional}

\subsection{The Oscillator Twisted Partition Function}
We now compute the oscillator partition function.

\mbni{Proposition II.1.} {\it Let $H=H_0$ and suppose that $|\gamma|<1$.  
Then 
\be
\fz_\gamma = \frac{1}{|1-\gamma|^2} 
  = \frac{e^{m\beta}}{4 |\sin(\frac{\w\th+im\beta }{2})|^2} > 0\;.
\ee{partitionfunction}
}
\mbni{Proof.}  We evaluate $\fz_\gamma$ by taking the trace 
in the basis $|n_+,n_-\rangle$  of simultaneous eigenstates of $N_\pm$,
introduced in \S II.1.  The result is that 
\be
\fz_\gamma 
  = \Tr_\fh\l(\uth^*  e^{-\beta H_0} \r) 
  = \sum_{n_+, n_- =0}^\infty \gamma^{n_+} \og^{n_-}
  = |1-\gamma|^{-2}\;.
\ee{EvaluateZ}

\subsection{The Pair Correlation Function and Time-Ordered Products}
Let $T$ denote an operator with domain $\D(T)$.
Define the imaginary time propagation $T(t)$ of $T$ by the general 
Hamiltonian $H$ to be 
\be
T(t) = e^{-tH} Te^{tH}\;.
\ee{imag}
For $t>0$, we obtain an operator with domain $e^{-tH}\D(T)$.  In
general $T(t)^\ast$ and $(T^\ast)(t)$ are equal only for
$t=0$.  Denote the time-ordered product of $T_1(t_1)\clips,T_n(t_n)$ by
$(T_1(t_1)\cdots T_n(t_n))_+$, with the definition 
\be
(T_1(t_1)\cdots T_n(t_n))_+= T_{i_1}(t_{i_1}) T_{i_2} (t_{i_2}) \cdots
T_{i_n}(t_{i_n})\; , \hensp{where} t_{i_1}\le t_{i_2} \le \cdots \le
t_{i_n}\;.
\ee{timeorder}

Define the {\it pair correlation function} by
\be
C_\gamma (t,s) = \lra{(\oz(t)z(s))_+}_{\gamma}\;.
\ee{defnc}
We now make a side technical assumption that $0\le H$ and that $z(H+I)^{-1}$
is bounded. This is a fact in case $H=H_0$.  We require this to use
cyclicity of the trace in order to establish the following twist condition.

\mbni{Proposition II.2.} {\it Suppose that $0\le H$, that $e^{-\beta H}$ 
has a trace,  and that $z(H+I)^{-1}$ is bounded. 
Then the pair correlation function $C_\gamma(t,s)$ of \refp{defnc} 
is a function of the difference coordinate $\xi=t-s \in [-\beta,\beta ] $,
and $C_\gamma(\xi)$ satisfies the twist relations
\be
C_\gamma(\xi+\beta) = e^{ -i\w\th}C_\gamma(\xi)\;, \hensp{if} \xi\le0 \;,
\qquad\hensp{and }\qquad
C_\gamma(\xi-\beta) = e^{ i\w\th}C_\gamma(\xi)\;, \hensp{if} \xi\ge0\;.
\ee{TwistRelation}
}
\vskip -18truept

\mbni{Proof.} We use the definition of the pair correlation
function and cyclicity of the trace. As $z(H+I)^{-1}$ is bounded, we
can also cyclically permute factors of $z$ and of $\oz$ in the heat-kernel
regularized trace.  Note that $\oz(t+t') = e^{-t'H} \oz(t) e^{t'H}$. 
Then for $0\le t\le s\le \beta$ and $0\le t+a\le s+a\le \beta$, we have 
\be
C_\gamma(t+a,s+a) = \lra{\oz(t+a)z(s+a)}_\gamma 
= \lra{ e^{-aH}\oz(t)z(s)e^{aH} }_\gamma =\lra{ \oz(t)z(s) }_\gamma
= C_\gamma(t,s)\;.
\ee{PCtranslation}
For $0\le s \le t \le \beta$ and $0\le s + a \le t +a \le \beta$ 
the same result holds. Hence $C_\gamma(t,s) = C_\gamma(\xi)$ is a 
function of the time difference.

We extablish the twist property in a similar fashion. For 
$0\le t\le s\le \beta$, we have 
\beq
C_\gamma(t-s+\beta)
&=& \lra{(\overline{z}(t- s +\beta) z(0))_+}_{\gamma }
= \lra{z(0) \overline{z}(t - s + \beta)}_{\gamma }\nn\\
&=& e^{ -i\w\th} \lra{\overline{z}(t - s + \beta) z(\beta) }_{\gamma }
= e^{ -i\w\th} C_\gamma(t-s)\;. 
\label{TRDerived1}
\eeq
For the case $0\le s\le t\le\beta$, write 
\beq
C_\gamma(t-s -\beta)
&=& \lra{(\overline{z}(t-s) z(\beta))_+}_{\gamma }
= \lra{\overline{ z}(t-s) z(\beta)}_{\gamma }\nn\\
&=& e^{ i\w\th}\lra{z(0) \overline{ z}(t-s)}_{\gamma }
= e^{ i\w\th} \lra{(\overline{z}(t-s) z(0))_+ }_{\gamma }\nn\\
&=& e^{ i\w\th} C_\gamma(t-s)\;. 
\label{TRDerived2}
\eeq

\subsection{Holonomy Moves}
In this brief section we explain the idea of {\it holonomy moves}, namely the
elementary steps in the holonomy expansion that 
we introduced in \cite{holonomy}.  We explore here an elementary special
case of a holonomy move. This case arises if we assume that there exists a 
complex-valued function $s=s(m,\beta, \th)$ such that $S$ 
satisfies the commutation relation, 
\be
S e^{-\beta H} \uth^* = s e^{-\beta H}\uth^* S \;.
\ee{srepresentation}
Let $\lra{ST}_\gamma$ denote the twisted expectation of the product of 
$S$ and $T$.  

An $S$-holonomy move is an identity that involves 
moving $S$ through the trace, and cyclically back to its original position.  
In particular, we may move $S$ in a clockwise fashion and exploit the
relation \refp{srepresentation} to permute $S$ cyclically in the trace.
We find that
\be
\lrag{ST} = \lrag{TS} +  \lrag{[S,T]} 
= s \lrag{ S T} + \lrag{[S,T]} \;.
\ee{Sholonomy1}
As long as $s \neq 1$, this $S${\it -holonomy move}
generates the identity between expectations, 
\be
\lrag{ST} = \frac{1}{(1-s)}\lrag{[S,T]} \;.
\ee{Sholonomy2}
The same identity can be written as 
\be
\lrag{ST} = \frac{-s^{-1}}{(1-s^{-1})} \lrag{[S,T]} \;.
\ee{Sholonomy4}
We might also interpret this latter identity as coming from counter-clockwise
movement of $S$ through the expectation, which generates the relation
\be
\lrag{ST} = s^{-1} \lrag{TS} 
= s^{-1}  \lrag{ST} 
      +  s^{-1}  \lrag{[T,S]}\;.
\ee{Sholonomy3}
We use the $S$-holonomy identity \refp{Sholonomy2}, \refp{Sholonomy4} below.

\subsection{The Twisted Oscillator Gibbs Functional is Gaussian}
In this section we show that the twisted Gibbs expectation is Gaussian. Our 
proof is an elementary application of two holonomy moves.  Let
$\{(t_1,s_{i_1}), (t_2,s_{i_2})\clips, (t_n,s_{i_n})\}$ denote one of the 
$n!$ pairing of $\{t_1\clips, t_n\}$ with $\{s_1\clips,s_n\}$.

\mbni{Proposition II.3.} {\it
Let $H=H_0$ and $|\gamma|<1$. Then the pair correlation function
$C_\gamma(t,s) = C_\gamma(\xi)$ equals 
\be
C_\gamma(\xi) 
= {1\over 2m} {\gamma\over (1- \gamma) } e^{-m\xi} 
  + {1\over 2m} {\og \over (1 - \og) } e^{m\xi}  
  + {1\over 2m} e^{-m|\xi|}\;,
\ee{kernel}
and 
\be
0\, < C_\gamma(0) = {1\over 2m} \lrp{ {1-|\gamma|^2\over |1-\gamma|^2}}\; . 
\ee{covequaltime}

Furthermore, the general
expectation \refp{qfunctional} of time-ordered products of coordinates 
satisfies the Gaussian relation
\be
\lra{\lrp{ \oz(t_1)\cdots \oz(t_n)z(s_1)\cdots z(s_r)}_+}_\gamma 
= \delta_{nr}
\sum_{\rm pairings} C_\gamma(t_1-s_{i_1}) \cdots C_\gamma(t_n-s_{i_n})\;.
\ee{moments}
}

\mbni{Remark.}
In particular, if $t_1=t_2= \cdots = t_n=s_1=\cdots = s_n$, then
$\lra{\,\lrp{\oz z}^n_{\phantom{+}}}_\gamma = n!\, C_\gamma(0)^n$.

\mbni{Proof.} Let us begin by verifying \refp{moments}. In the course of this
we also show \refp{kernel}. Hence \refp{covequaltime} results from 
the evaluation 
\be
C_\gamma(0) =
{1\over 2m} \lrp{ {\og\over 1-\og} 
+{\gamma\over 1-\gamma}+1} = {1\over 2m} 
\lrp{{1-|\gamma|^2\over |1-\gamma|^2}} \; .
\ee{pair}
We argue that \refp{moments} vanishes except for $n=r$. 
As $\uth$ commutes with $H$, the expectation
$\lra{\,\cdot\,}_\gamma$ is invariant under $\uth$ in the sense that
\be
\lra{T}_\gamma= \lra{\uth T\uth^\ast}_\gamma\; . 
\ee{invariance}
We take for $T$ the time-ordered product,
$T = (\oz(t_1)\cdots \oz(t_n)z(s_1)\cdots z(s_r))_+$. Then we have
\be
\uth T\uth^\ast =e^{i\w\th(r-n)}T\;,
\ee{transform}
and
\be
\lra{T}_\gamma = e^{i\w\th(r-n)}\lra{T}_\gamma 
\ee{noninvariance}
for all $\th$.  In case $n\ne r$, this can only be the case for
$\lra{T}_\gamma=0$.  Thus we restrict attention to the case $n=r$.

It is no loss of generality to relabel the times so that the 
smallest time is either $s_1$ or $t_1$.  Let us first consider the case
that $0\le s_1\le t_1\le \beta$. Define
\be
T = e^{s_1 H}\lrp{\oz(t_1) \cdots \oz(t_n) z(s_2)\cdots z(s_n)}_+ 
               e^{-s_1 H}\;.
\ee{Tdefinition}
Furthermore, let $T_i$, for $1\le i\le n$,  denote $T$ with the factor 
$\oz(t_i)$ omitted.  We rewrite the expectation \refp{moments}
in terms of the operators $T$ and $T_j$. Since $H$ and $\uth$ commute, 
by cyclicity of the trace we can write  
\be
\lrag{ \lrp{ \oz(t_1)\cdots \oz(t_n) z(s_1)\cdots z(s_n)}_+ }
= \lrag{z T}\;.
\ee{expandexpectation1}
We claim that this expectation satisfies the 
{\it Gaussian recursion relation}
\be
\lra{\lrp{ \oz(t_1)\cdots \oz(t_n)z(s_1)\cdots z(s_r)}_+}_\gamma 
= \lra{z T}_\gamma 
= \sumj C_\gamma(t_j-s_1)\lra{T_j}_\gamma\;.
\ee{zrecursion}
The second case, on the other hand, is $0\le t_1\le s_1\le \beta$.
In this case we define
\be
S = e^{t_1 H}\lrp{\oz(t_2) \cdots \oz(t_n) z(s_1) \cdots z(s_n)}_+ 
               e^{-t_1 H}\;,
\ee{Sdefinition}
and we have 
\be
\lrag{ \lrp{ \oz(t_1)\cdots \oz(t_n) z(s_1)\cdots z(s_n)}_+ } 
= \lrag{\oz S}
\ee{expandzerexpectation1}
Define $S_i$, for $1\le i\le n$,  as the operator $S$ with the factor 
$z(s_i)$ omitted.  
In this case, we claim that this expectation satisfies the 
{\it conjugate Gaussian recursion relation}
\be
\lrag{ \lrp{ \oz(t_1)\cdots \oz(t_n) z(s_1)\cdots z(s_n)}_+ }
= \lrag{\oz S}
= \sumj C_\gamma( t_1 - s_j)\lra{S_j}_\gamma\;.
\ee{zbarrecursion}
The identity \refp{moments} then follows by iteration of the
recursion relations \refp{zrecursion} and 
\refp{zbarrecursion}.

We now prove \refp{zrecursion} and \refp{zbarrecursion}, and thereby 
complete the proof of the proposition.  Begin the proof of \refp{zrecursion} 
by rewriting \refp{expandexpectation1} in a form where we 
can apply a holonomy move \refp{Sholonomy2}, with $S$ equal to an 
annihilation operator or a creation operator. Using \refp{coord}, 
\be
\lrag{\lrp{ \oz(t_1)\cdots \oz(t_n)z(s_1)\cdots z(s_n)}_+  }
= \frac{1}{\sqrt{2m}}\lrag{(a_+^* + a_-) T}\;.
\ee{expandexpectation2}
Let us consider the two terms separately. 

In performing a holonomy move, we need to evaluate the commutators
between $a_-$ and $T$ or between $a_+^*$ and $T$.  One can commute 
$a_\pm$ with $z$ or $\oz$ using
\be
[a_-,\oz]=(2m)^{-1/2}=-[a_+^\ast,\oz]
\qquad\hensp{and}\qquad
[a_+^\ast ,z] = [a_-,z]=0\;.
\ee{elemcomm} 
These identities follow from \refp{coord} and the canonical commutation 
relations \refp{ccr}.  In order to commute $a_\pm$ with $e^{-tH}$ or $\uth$, 
use the basic identities that are a consequence of \refp{ham},
\be
a_\pm e^{-tH} = e^{-tm}e^{-tH} a_\pm\;, 
\qquad\hensp{and} \qquad
a_\pm \uth^* =e^{\mp i\w\th}\uth^*  a_\pm\;. 
\ee{atrans}
Thus  
\be
a_- e^{-\beta H}\uth^*  = \gamma e^{-\beta H}\uth^*  a_- \;,
\qquad\hensp{and} \qquad
a_+ e^{-\beta H}\uth^*  = \og e^{-\beta H}\uth^*  a_+ \;,
\ee{aminus}
The relation \refp{aminus} shows that $a_-$ and $a_+$ satisfy the hypothesis
\refp{srepresentation} with $s=e^{-m\beta + i\w\theta} = \gamma$ and 
$s=\og$ respectively.  Furthermore, by substituting $U(-\th)$ for $\uth$
and taking adjoints, we obtain similar relations for $a_\pm^*$,
\be
a_+^* e^{-\beta H}\uth^*  = \gamma^{-1} e^{-\beta H}\uth^*  a_+^* \;,
\qquad\hensp{and} \qquad
a_-^* e^{-\beta H}\uth^*  = \og^{-1} e^{-\beta H}\uth^*  a_-^* \;.
\ee{aplus} 
Hence $a_\pm$ and their adjoints may be used to 
define holonomy moves of the form \refp{Sholonomy2} or \refp{Sholonomy4}.

Let us begin by expanding the second term in \refp{expandexpectation2}.  
We perform an $a_-$-holonomy move and substitute the identity 
\refp{Sholonomy2}. Using \refp{elemcomm} and \refp{atrans} we have
\be
[a_-, T] = \sumj \frac{1}{\sqrt{2m}} e^{-m(t_j - s_1)} T_j\;.
\ee{tcommutator}
Therefore 
\be 
{1\over \sqrt{2m}} \lra{a_{-} T }_\gamma  
     =  {1\over 2m}  {1\over (1-\gamma)}
\sumj e^{-m(t_j-s_1)} \lra{T_j}_\gamma\;. 
\ee{agholo}
On the other hand, for the first term in \refp{expandexpectation2}, 
we choose perform an $a_+^\ast$-holonomy move.  In this case, we refer to  
\refp{aplus} that gives the representation \refp{srepresentation} with 
$s^{-1}=\og$. Use \refp{elemcomm} to obtain
\be
[a_+^\ast, T] = - \sumj \frac{1}{\sqrt{2m}} e^{m(t_j - s_1)} T_j\;.
\ee{tpluscommutator}
Then we write the alternative form \refp{Sholonomy4} of the $a_+$-holonomy 
move as 
\be
{ 1\over\sqrt{2m} } \lra{a^\ast_+ T}_\gamma 
= {1\over 2m}  {\og \over (1-\og)} \sumj e^{m(t_j-s_1)} \lra{T_j}_\gamma\;. 
\ee{adgholo}
Add \refp{agholo} and \refp{adgholo}, so that 
\refp{expandexpectation1}---\refp{expandexpectation2} become
\be
\lrag{z T} 
= \sumj {1\over 2m} \l( { \og\over (1-\og) } e^{m(t_j-s_1)} 
+ {1\over (1- \gamma)}  e^{-m(t_j-s_1)}\r) \lra{T_j}_\gamma \;.
\ee{addrecursion}
Note that in the case $n=1$, then $T_1 = I$ and 
$\lra{zT}_\gamma = C_\gamma(t_1-s_1)$.  Thus we have also proved that 
(in the case that $0\le s_1 \le t_1\le \beta$) 
\be
C_\gamma(t_1-s_1) 
= {1\over 2m} \l( { \og\over (1-\og)} e^{m(t_1-s_1)} 
+ {1\over (1-\gamma)}  e^{-m(t_1-s_1)}\r) \;,
\ee{GetPairCorrelation}
as claimed in \refp{kernel}.
Insert the identity \refp{GetPairCorrelation} into 
\refp{addrecursion}, to yield the desired recursion relation
\refp{zrecursion}. 

Next, let us treat the case $0\le t_1 \le s_1\le \beta$ and establish 
\refp{zbarrecursion}.  In place of \refp{expandexpectation2}, we use
\be
\lrag{\lrp{ \oz(t_1)\cdots \oz(t_n)z(s_1)\cdots z(s_n)}_+  }
= \frac{1}{\sqrt{2m}}\lrag{(a_+ + a_-^*) S}\;,
\ee{expandexpectation3}
with $S$ given by \refp{Sdefinition}.  Consider first the $a_+$-term, and
perform an $a_+$-holonony; observe that \refp{aplus} ensures $s=\og$.  
Furthermore, in place of \refp{tcommutator}, we use
\be
[a_+, S] = \sumj \frac{1}{\sqrt{2m}} e^{-m(s_j - t_1)} S_j\;.
\ee{scommutator}
Thus
\be
\frac{1}{\sqrt{2m}} \lrag{a_+ S} 
= \frac{1}{2m} \frac{1}{(1-\og)} \sumj e^{-m(s_j - t_1)} \lrag{S_j} \;.
\ee{expandexpectation4}
Similarly, we obtain
\be
\frac{1}{\sqrt{2m}} \lrag{a_-^* S} 
= \frac{1}{2m} \frac{\gamma}{(1-\gamma)} 
   \sumj e^{m(s_j - t_1)} \lrag{S_j} \;.
\ee{expandexpectation5}
Adding \refp{expandexpectation4} and \refp{expandexpectation5} we infer,
\be
\lrag{\oz S} 
= \sumj \frac{1}{2m} \l( \frac{1}{(1-\og)} e^{-m(s_j - t_1)} 
 + \frac{\gamma}{(1-\gamma)} e^{m(s_j - t_1)} \r) \lrag{S_j} \;.
\ee{addconjrecursion}
In the case $n=1$, note that $S_1 = I$, so that \refp{addconjrecursion}
reduces to 
\be
C_\gamma(t_1-s_1) 
  = \frac{1}{2m} \l( \frac{1}{(1-\og)} e^{-m(s_1 - t_1)} 
      + \frac{\gamma}{(1-\gamma)} e^{m(s_1 - t_1)} \r) \;,
\ee{GetConjPairCorrelation}
thus establishing \refp{kernel} in the case $0\le t_1\le s_1\le \beta$.
Inserting this identity into \refp{addconjrecursion} 
completes the proof of \refp{zbarrecursion} and of the proposition.

\subsection{Fourier Expansions}
We encouter functions $f(t)$ on the interval $[0,\beta]$ that are continuous 
and satisfy the twisted periodicity condition 
\be
f(\beta) = e^{ -i\w\th} f(0)\;.
\ee{FunctionTwistBC}
According to Proposition II.2, the kernels $C_\gamma(t)$ and 
$C_\gamma(t-\beta)$ have this property. We expand such functions in Fourier 
representation of the form
\be
f(t) = \sum_{E\in \frac{2\pi}{\beta}\Z} 
       \hat f(E) e^{iEt} e^{ -i\w\th t/\beta}\;.
\ee{FunctionFourier}
In fact, on the interval $[0,\beta]$, linear combinations of the Fourier 
basis functions $\l\{e^{ iEt}\r\}$ with $\beta E\in 2\pi\Z$ are a dense
subset of $L^2([0,\beta], dt)$.  Furthermore the smooth function of $t$,
$e^{ -i\w\th t/\beta}$, defines a multiplication operator on 
$L^2([0,\beta])$. The modulus of this function is one, so the multiplication
operator is unitary.  Thus the functions 
$\l\{e^{iEt} e^{ -i\w\th t/\beta} \r\}$ also span $L^2$. Functions $f(t)$ 
on $[0,\beta]$, that have pointwise convergent Fourier 
representations \refp{FunctionFourier}, both satisfy the boundary condition
\refp{FunctionTwistBC} also extend (using this representation) to functions 
on the line $\R$ that satisfy the twist relation
\be
f(t+\beta) = e^{ -i\w\th} f(t)\;.
\ee{FunctionTwistRel}
We may also rewrite this relation as follows. Define the set 
\be
K_\gamma= \l\{E: \beta E= 2\pi \Z - \w\th \r\}\;. 
\ee{kdef}
Then 
\be
f(t) = \sum_{E\in K_\gamma} \hat f(E) e^{iEt}\; , 
\ee{fourier}
with
\be
\beta\hat f(E) = \int^\beta_0 f(t)e^{-iEt}dt\; , 
\ee{coeff}
and
\be
\int^\beta_0 |f(t)|^2 dt = \beta \sum_{E\in K_\gamma} |\hat f(E)|^2 \; .
\ee{unitarity}

\subsection{The Oscillator Pair Correlation Operator}
Regard the pair correlation function $C_\gamma(t,s)$ as the integral kernel 
of an operator $C_\gamma$ on $L^2([0,\beta], dt)$, so
\be
(C_\gamma f)(t) = \int_0^\beta C_\gamma(t,s) f(s) ds\;.
\ee{PCop}
Denote $C_\gamma$ the {\it pair correlation operator}. 

\mbni{Proposition II.4} {\it 
The operator $C_\gamma$ is self-adjoint, strictly positive, and 
bounded. In fact, 
\be
0\, < \,C_\gamma = C^\ast_\gamma \, \le \, M\; , 
\ee{sa}
where the upper bound $M$ equals
\be
M= \max\l\{ \lrp{ \lrp{ {\w\th\over \beta}}^2+m^2}^{-1}\;,
\lrp{\lrp{ {2\pi - \w\th\over \beta}}^2 +m^2}^{-1} \r\}\;.
\ee{bound}
In particular, $M$ is singular only if $\th\in \{0,2\pi/\w\} = \sing$
and also $m \to 0$.
}

\mbni{Proof.}
The function \refp{kernel} satisfies
\be
C_\gamma(s,t)=C_{\og} (t,s)=\overline{C_\gamma(t,s)}\; . 
\ee{hermitian}
Thus $C_\gamma$ is a hermitian operator.  Furthermore, $C_\gamma(t,s)$ 
is a bounded
function on the compact set $[0,\beta]\times [0,\beta]$, so the operator
$C_\gamma$ is Hilbert--Schmidt.  In particular $C_\gamma$ is bounded and
self-adjoint.  In order to show that $C_\gamma$ is strictly positive and to
compute the upper bound $M$, we compute the spectrum of $C_\gamma$.  

We compute the Fourier series for $C_\gamma(\xi)$.  
The variable $\xi$ lies in the interval $[-\beta,\beta]$ and the 
$C_\gamma(\xi)$ satisfies the twist relation \refp{TwistRelation}.  Thus 
\be
C_\gamma(\xi) = \sum_{E\in K_\gamma} \hat{C}_\gamma (E) e^{iE\xi} 
\ee{cspectrum}
with the consequence that
\beq
\lra{f,C_\gamma g} 
&=& \int^\beta_0 dt \int^\beta_0 ds \overline{f(t)} C_\gamma(t-s)g(s)\nn\\
&=& \beta^2 \sum_{E\in K_\gamma} \hat{C}_\gamma (E) 
      \overline{\hat f(E)} \hat g(E) \; . 
\label{csexpectation}
\eeq
This identity gives a diagonalization of $C_\gamma$, so the 
spectrum of $C_\gamma$ is the set of values of 
\be
\beta \hat C_\gamma (E) = \int^\beta_0 C_\gamma(\xi) e^{-iE\xi}d\xi \;,
\ee{fourierc}
as $E$ ranges over $K_\gamma$, and for such $E$ we have
\be
\frac{1}{1-\gamma} \int_0^\beta e^{-(m+iE)\xi}d\xi = {1\over m+iE}\;,
\qquad\hensp{and}\qquad
{\og\over 1-\og} \int^\beta_0 e^{(m-iE)\xi} d\xi = {1\over m-iE}
 \;.
\ee{ftexp} 
Hence we infer from \refp{kernel} that
\be
\beta \hat C_\gamma(E) 
= {1\over 2m} \lrp{ {1\over m+iE}+{1\over m-iE}} 
= {1\over E^2 +m^2}\;.
\ee{answer}
The positivity of $C_\gamma$ follows, as well as the value of
$M=\Vrt{C_\gamma}$ in \refp{bound}, namely
\be
M=\sup_{E\in K_\gamma} \lrp{E^2 +m^2}^{-1}\;.
\ee{Mvalue}
Hence the proof of the proposition is complete.

We summarize the last calculation by the statement,
\mbni{Proposition II.5} {\it The pair correlation function $C_\gamma(t,s)$
has a Fourier series}
\be
C_\gamma(t,s) = \frac1\beta \sum_{E\in\K_\gamma} 
\frac1{E^2 + m^2} e^{iE(t-s)}\;,
\ee{PCFourierQuantum}
where $K_\gamma$ is defined in \refp{kdef}.

\subsection{The Oscillator Pair Correlation Function Has a Complex Period}
The pair correlation function 
$C_\gamma(t,s) = C_\gamma(t-s)$ is a
function of the difference variable $\xi=t-s$.  We now see that
$C_\gamma(\xi)$ has a natural extension to all complex 
$\xi$, and this function has a complex period. 
Let $\Re(\xi)$ and $\Im(\xi)$ denote the real and
imaginary parts of $\xi$ respectively.  This  
extension is neither holomorphic nor antiholomorphic, but in each strip
$n\beta < \Re(\xi) < (n+1)\beta$, it is the sum of a holomorphic and an 
antiholomorphic part.  Furthermore, it is continuous, single-valued, and 
periodic, with a complex period $\eta$.  The extension also obeys the 
reflection principle
\be
{C_\gamma(\xi)} = \overline{C_\gamma(-\xi)} \;. 
\ee{reflection}

For $\xi$ real, the limiting case $\gamma=0$ (obtained from 
$C_\gamma(\xi)$ as 
$\beta\to\infty$) equals ${1\over 2m}e^{-m|\xi|}$.  Let us define the 
extension of $C_0(\xi)$ to all $\xi\in \C$ by 
\be
\FC_0(\xi ) = \l\{ 
\begin{array}{ll} 
\frac{1}{2m} e^{m\xi} &\ \hensp{if}\  {\Re}(\xi)\le 0\\
\frac{1}{2m} e^{-m\overline{\xi}} &\ \hensp{if}\ 0 \le {\Re}(\xi)\;. 
\end{array}\r. 
\ee{czero}
Clearly, \refp{czero} satisfies \refp{reflection}.  
Introduce the complex periodic function
\be
\FC_\gamma(\xi)= \sum_{k\in \Z} \FC_0(\xi + k\eta)\;, 
\ee{cperdef}
defined for all complex $\xi$. We choose the period 
$\eta$ of $\FC_\gamma(\xi)$ to equal 
\be
\eta=\beta + i{\w\th\over m}\; , \qquad \hensp{so} \qquad 
\gamma=e^{-m\eta}\;.
\ee{cperiod}
We now show that the sum \refp{cperdef} agrees with the function 
\refp{kernel} in the domain of definition of the latter, and thereby 
defines a natural extension of \refp{kernel} to the entire complex 
$\xi$ plane.

\mbni{Proposition II.6.} {\it Let $\FC_\gamma(\xi)$ and $\eta$ be given by 
\refp{cperdef} and \refp{cperiod} respectively, with $|\gamma|<1$. Then}
\begin{itemize}
\ritem{a.} Given $n \in \Z$, let $\xi$ lie in the strip 
$n\beta\le {\Re}(\xi)\le (n+1)\beta$. We also write $\xi = n\beta+\xi_1$.  
Then
\be
\FC_\gamma(\xi) = \frac{e^{ -i n\w\th}}{2m} 
\l[ e^{-m\overline{\xi_1}} \l(\frac{1}{1-\gamma}\r)
 + e^{m\xi_1} \l( \frac{ \og} { 1-\og} \r) \r] \; . 
\ee{qgeneral}

\ritem{b.} {\it For real $\xi$ in the interval $-\beta\le \xi\le \beta$, 
the function $\FC_\gamma(\xi)$ in \refp{qgeneral} agrees with \refp{kernel}. 

\ritem{c.} The function $\FC_\gamma(\xi)$ in \refp{qgeneral} obeys 
reflection symmetry \refp{reflection}, and also satisfies the periodicity 
relations
\be
\FC_\gamma(\xi+\eta) = \FC_\gamma(\xi)\;,\hensp{and}
\FC_\gamma(\xi+\beta)=e^{ -i\w\th}\FC_\gamma(\xi)\; , 
\ee{relations}
for all $\xi\in \C$. }\end{itemize}

\mbni{Proof.} The representation \refp{qgeneral} follows from \refp{czero}
and the definition \refp{cperdef}.  We evaluate the sum over translations 
$k\eta$,  $k\in\Z$, by splitting the sum into the two ranges 
$-\infty < k \le -n-1$ and $-n \le k < \infty$.  
Note that the expression \refp{qgeneral} is single valued.  
In fact, it is the convergent sum of translations of the single-valued
function \refp{czero}. The value of \refp{qgeneral} with $\Re(\xi)=n\beta$ is
\be
\frac{e^{-in\w\th}} {2m} e^{im {\Im}(\xi_1) }
\l(\frac{\gamma}{1-\gamma} +\frac{1}{1-\og}\r)
= \frac{e^{ -in\w\th}} {2m} e^{im {\Im}(\xi_1) }
\l( \frac{1-|\gamma|^2}{|1-\gamma|^2} \r)\;.
\ee{righthand}
This could be obtained either from the formula \refp{qgeneral} applied with 
$\xi_1$ purely imaginary, or with $\Re(\xi_1)=\beta$. Thus we have verified
part (a) of the proposition.

The proof of statement (b) of the proposition merely involves the 
inspection of the function defined by \refp{kernel}. Compare it with
\refp{qgeneral} in the case that $-\beta\le\xi\le\beta$ (namely for 
$n=-1$ yielding $-\beta\le\xi\le0$ and for $n=0$ yielding
$0\le\xi\le\beta$).  We easily conclude that the two functions agree.  

In order to verify the reflection symmetry \refp{reflection}, we use this 
symmetry for $\FC_0$. Then  
\be
\FC_\gamma(-\xi) = \sum_{n\in\Z} \FC_0(-\xi + n\eta) 
= \sum_{n\in\Z} \FC_0(-\xi - n\eta) 
= \sum_{n\in\Z} \overline{\FC_0(\xi + n\eta)} =\overline{\FC_\gamma(\xi)}\;,
\ee{reflectiongamma}
establishing the desired identity. 
Also the fact that $\FC_\gamma(\xi)$ is
periodic with period $\eta$ is a consequence of the definition
\refp{cperdef}.  Then, for all $\xi$,
\be
\FC_\gamma(\xi+\beta)= \FC_\gamma(\xi+\beta-\eta) 
= \FC_\gamma\lrp{\xi - i{\w\th\over m}}\;.
\ee{twoperiods}
But $\xi$ and $\xi -i {\w\th\over m}$ have the same real part.  Hence
using \refp{qgeneral} to evaluate
\refp{twoperiods}, we find that translation by $-i\w\th/m$ does not affect
$n$. In only affects the exponential factors in \refp{qgeneral}, and each
is multiplied by $e^{ -i\w\th}$. This 
ensures that $\FC_\gamma( \xi + \beta)=e^{ -i\w\th} \FC_\gamma ( \xi)$. 
This verifies statement (c). We have now checked all parts of 
Proposition II.6.

\section{The Gaussian Path-Space Measure $\dmug$}
\setcounter{equation}{0}
\subsection{The Twisted Laplacian and its Green's Operator}
Let $\st([0,\beta])$ denote the space of $C^\infty$ functions 
on $[0,\beta]$ that have Fourier representations of the form
\be
f(t) = \sum_{E\in K_\gamma} \hat f(E) e^{i(E - \w\th/\beta)t}\;,
\ee{FourierFunctionOft}
where the coefficients $\hat f(E)$ are rapidly decreasing in $E$.  Such 
functions are $C^\infty$ and satisfy the twist relation 
\be
f(t+\beta) = e^{ -i\w\th} f(t)\;.
\ee{TwistForFunctions}
Endow $\st([0,\beta])$ with the usual Fr\^echet topology, given by the 
countable set of norms
\be
||f||_n = \sup_{E\in \frac{2\pi}{\beta}\Z} (1+E^2)^n |\hat f(E)|\;,
\ee{norms}
and as such $\st([0,\beta])$ is a nuclear space.
The operator $\frac{d}{dt}$ acting on $L^2([0,\beta])$ with domain 
$\st([0,\beta])$ is skew-symmetric.  Let $D_\th$ denote its closure.
The operator $-iD_\th$ is self-adjoint.\footnote{Note that $\frac{d}{dt}$ 
defined on the domain $C^\infty_0([0,\beta])$ of smooth, compactly 
supported functions, has deficiency indices $(1,1)$, so $\frac{d}{dt}$  
has a one-parameter family of skew-adjoint extensions.  Here $\th$ 
parameterizes this family, as Wightman discussed in his 
1964 Carg\`ese lectures \cite{Wightman}.} 
Also the twisted Laplace operator $D_\th^2 $ is the self-adjoint closure of 
$\frac{d^2}{dt^2}$ on the domain $\st([0,\beta])$. 
The resolvent of the twisted Laplace operator is called the twisted Green's 
operator. We designate it 
\be
\green  = (-D_\th^2 +m^2)^{-1}\;,
\ee{twistedLaplace}
and call $m>0$ the {\it mass} using the usual name from physics.
As an operator on $\st([0,\beta])$, or as an operator on $L^2([0,\beta])$,
the twisted Green's operator $\green $ has an integral kernel 
$\green(t,s)$ that we denote the twisted Green's function.  The 
twisted Laplace operator is translation invariant, and hence so is 
the twisted Green's operator.  In terms of the Green's function, 
$\green (t,s)=\green (t-s)$.

\mbni{Theorem III.1.} {\it The Green's operator $\green$ equals the 
pair correlation operator $C_\gamma$ defined in \refp{PCop},}
\be
\green = C_\gamma = \lrp{-D^2_\th+m^2}^{-1} \;.
\ee{periodical}

\mbni{Proof.}
Identifying the operators $C_\gamma$ and $\green$ is equivalent to 
identifying their integral kernels, or the Fourier series of these kernels.  
We computed the Fourier series for the kernel of $C_\gamma$ in 
Proposition II.5, and clearly this 
agrees with the Fourier series for the kernel of $G_\theta$. Thus
the operators agree.  
In addition, the orthonormal basis of functions 
\be
e_E(t) = \beta^{-1/2}e^{iEt}\in \st([0,\beta])\;, \quad E\in K_\gamma\;,
\ee{eigenfn}
are eigenfunctions of the operator $C_\gamma$, corresponding to
eigenvalues are $(E^2+m^2)^{-1}$.   

\subsection{The Measure}
Consider the Schwartz space $\st$ with the topology given by the countable
set of norms \refp{norms}. 
Let $\wp\in\spt$ denote a path in the space of distributions dual 
to $\st$.  Continuous functions $\wp(t)$ pair with elements of $\st$ 
through the bilinear relation 
\be
\wp(f) = \int^\beta_0 \wp(t) f(t)dt\;,
\ee{pairing}
and this pairing extends by continuity to $\spt$.
The pair correlation operator $C_\gamma$ maps $\spt$ continuously into 
itself. Therefore there is an adjoint map $C_\gamma^+$ that is a continuous
linear transformation of $\spt$ into itself, defined by
\be
(C_\gamma^+\wp)(f) = \wp(C_\gamma f)
\ee{adjointC}
The kernel $C_\gamma^+(t,s) = C_\gamma^+(t-s)$ of $C_\gamma^+$ is 
the complex conjugate of the kernel of $C_\gamma$, so using the twist relation
of Proposition II.2, we infer 
\be
C_\gamma^+(t-s+\beta) = e^{ i\w\th} C_\gamma^+ (t-s) \;.
\ee{AdjointTwistRelation}
As a consequence, continuous elements $\wp$ satisfy a twist relation dual to
the twist relation \refp{FunctionTwistRel} for functions, namely 
\be
\wp(t+\beta) = e^{ i\w\th} \wp(t)\;.
\ee{dualbc}

The identification of $C_\gamma$ in Theorem III.1 ensures that if 
\be
|\gamma| \le 1\;,\quad \gamma\ne 1\;,
\ee{qcondition}
then $C_\gamma$ is a continuous map of $\st$ to $\st$ in the 
Schwartz space topology. 
Furthermore, acting on $L^2([0,\beta])$, the operator 
$C_\gamma$ is strictly positive and has a bounded operator norm. 
Such a $C_\gamma$ is an appropriate covariance operator for 
a Gaussian probability measure on the dual space $\spt$ of generalized 
functions. See for example Chapter IV of \cite{Gelfand}, 
\S A.3--A.6 of \cite{QP}.

\mbni{Definition III.2.} {\it Let $|\gamma|\le 1$ and $\gamma\ne 1$.  
Let $\dmug=\dmug(\wp(\cdot))$ denote the Gaussian probability measure on 
$\spt$ with vanishing first moments
\be
\int \wp(f) \dmug = 0 = \int \overline{\wp(f)} \dmug\;,
\ee{firstmoments}
and with second moments
\be
\int \wp(f)\wp(g) \dmug =  0 \; , 
\hensp{\it and} 
\int \overline{\wp(f)} \wp(g) \dmug = \lra{f,C_\gamma f}_{L^2}\;.
\ee{secondmoments}
}
The Gaussian recursion relation for moments of 
$\dmug $, along with \refp{secondmoments} ensure that the moments 
\be
\int_{\cs'_\th} \overline{\wp(t_1) \wp(t_2) \cdots \wp(t_n)} 
        \wp(s_1) \wp(s_2)\cdots \wp(s_r) \dmug  
\ee{nondiagonal}
vanish unless $n=r$.  As usual, the measure $\dmug $ is concentrated on
the set of H\"older-continuous functions on $[0,\beta]$ with exponent less 
than $\frac12$. 
We now identity that the twisted Gibbs functional $\lra{\,\cdot\,}_\gamma$ 
of \refp{qfunctional}, applied to coordinates, equals the moments of the 
measure $\dmug $.

\mbni{Proposition III.3. (Gaussian Feynman-Kac Identity)} {\it 
Let $H=H_0$ and $|\gamma|<1$. Consider the expectation 
\be
\lra{\,\cdot\,}_{\gamma} 
   = \frac{\Tr_{\fh} (U(\theta)^*\, \cdot \, e^{-\beta H_0})}
    {\Tr_{\fh} (U(\theta)^* e^{-\beta H_0})} \;.
\ee{qfunctional2}
Then expectations of time-ordered products of coordinates, 
defined in \refp{timeorder}, are moments of $\dmug $.}
\beq
&& \lra{ \lrp{ \oz(t_1)\oz(t_2)\cdots\oz(t_n) z(s_1)z(s_2)\cdots
z(s_r)}_+}_\gamma \nn\\
&& \hskip 2.3in
= \int_{\cs'_\th} \overline{\wp(t_1) \wp(t_2) \cdots \wp(t_n)} 
        \wp(s_1) \wp(s_2)\cdots \wp(s_r) \dmug  \;.\nn\\
\label{fkidentity}
\eeq
\mbni{Proof.} According to Proposition II.3, the left side of 
\refp{fkidentity} is a Gaussian functional, and it vanishes unless $n=r$.
The right side of \refp{fkidentity} is Gaussian by definition.   
Likewise in Definition III.2 the first moments of the
measure $\dmug (\wp(\cdot))$ are defined to vanish. 
Hence it is sufficient to identify the $n=r=1$ expectation on the left 
side of \refp{fkidentity} with the corresponding second moment
on the right side of \refp{fkidentity}. But the $n=1$ expectations agree
by definition, so the proof is complete.

\mbni{Remark 1.}
The twisted expectation \refp{fkidentity} of time-ordered products of 
coordinates agrees with the moments 
of the measure $\dmug $. So it is natural to use the same notation 
$\lra{\, \cdot\, }_\gamma$ for an integral in $\spt$ 
and also for the twisted Gibbs functional on the Hilbert space $L^2(\C)$.
This notation should cause no confusion, so we write 
\be
\lraug{F} = \int_{\spt} F(\wp, \ow) \dmug \;.
\ee{integral}

\mbni{Remark 2.}
The measures $d\mu$ are associated with distinct notions of positivity. 
The first notion is ordinary {\it positivity as a measure} stating that 
the integral of a positive function is positive,
\be 
\int |A|^2 d\mu_{\beta,\th} \ge 0\;. 
\ee{posmeasure} 
A second notion of positivity expresses the fact that the Gibbs state
at zero twist must be positive when evaluated on a
positive operator,
\be 
\lra{\hat A^*\hat A}_{\beta,0} \ge 0\;. 
\ee{functionalpos} 
This condition is called {\it reflection positivity} when it is 
expressed in terms of the measure.
Suppose that the operator $\hat A$ denotes a time-ordered product 
of the coordinates $z(t_j)$ and the $\oz(t_j)$'s, where 
$\beta/2 \le t_j\le\beta$.  Let $A$ denote the same product of
$\wp(t_j)$'s and $\overline{\wp(t_j)}$'s.  Define a linear time reflection 
operator $\Theta$ on paths so that it reflects the time of a path about 
the midpoint of the time interval $[0,\beta]$. In particular, let 
$(\Theta \wp)(t) = \wp(\beta-t)$. Extend this definition to  
functions of paths as $(\Theta A)(\wp) = A(\Theta\wp)$. 
Then \refp{functionalpos} is equivalent to the statement that 
\be 
\lra{\hat A^*\hat A}_{\beta,0} 
= \int \overline{(\Theta A)(\wp)} A(\wp) d\mu_{\beta,0} \ge 0\;,
\ee{reflectionpos} 
where $\overline{A(\wp)}$ denotes complex conjugation of the function
$A(\wp)$. For $\th=0$, the measure $d\mu_{\beta,0}(\wp_0)$ satisfies 
reflection positivity.  The reflection positivity of a Gaussian measure 
is equivalent to reflection positivity of its covariance.\footnote 
{For a discussion of reflection
positivity in more detail, see \cite{QP}.}  
 
\subsection{Gaussian Integration by Parts and Mass Renormalization}
The measure $\dmug$ with covariance $\cmug$ satisfies an integration by 
parts formula.  Let us consider the path $\wp$ and the complex conjugate path 
$\ow$ as varying independently, and let $F(\wp,\ow)$ denote a functional on 
on $\spt$.  We say that $F$ is differentiable if for $0<t<\beta$ the 
limits
\be
\frac{\part F}{\part\wp(t)} = 
\lim_{\ep\to0} \frac{F(\wp +\ep\delta_t,\ow)-F(\wp, \ow)}{\ep}
\quad\hensp{and}\quad
\frac{\part F}{\part\ow(t)} = 
\lim_{\ep\to0} \frac{F(\wp, \ow +\ep\delta_t)-F(\wp, \ow)}{\ep}
\ee{functionalderivative}
exist, where $\delta_t$ denotes the Dirac measure translated by $t$. 
We may assume that the limit defines an element of $\spt$.  Since
$C_\gamma$ acts on $\st$, it also acts on $\spt$, and we denote this action
by 
\be
\wp(\cdot)\to (C_\gamma\wp)(\cdot)=\int_0^\beta C_\gamma(\cdot, t)\wp(t)dt\;.
\ee{covarianceaction}

\mbni{Proposition III.4. (Integration by Parts)} {\it Let $F$ denote 
a differentiable, polynomially bounded function, and let $|\gamma|<1$. Then}
\be
\lraug{\wp(s) F} = 
\int_0^\beta \cmgu \lraug{\frac{\part F}{\part\ow(t)}}dt\;,
\quad\hensp{and}\quad
\lraug{\ow(t) F} = 
\int_0^\beta \cmgu \lraug{\frac{\part F}{\part\wp(s)}}ds\;.
\ee{ibps}

\mbni{Remark.} 
These identities follow immediately from the properties 
of Gaussian integrals.  The two relations \refp{ibps} are related 
to each other by complex conjugation.  
The continuity of the expectations
$\lraug{\frac{\part F}{\part\ow(t)}}$, etc., 
at the endpoints of the interval $s\in[0,\beta]$ ensures that the 
integration can be extended to the endpoint. This is the case if
$\wp$ is averaged with smooth test functions. 
In terms of the Feynman-Kac expectations, a special case integration
by parts corresponds to the recursion relation \refp{zrecursion} for the 
trace functional.  In this case we require Dirac measures as test functions.
Take, 
$F(\wp, \ow) = \wp(\delta_{s_2 - s_1}) \cdots \wp(\delta_{s_n - s_1})
\ow(\delta_{t_1 - s_1}) \ow(\delta_{t_n - s_1})$. Then 
\be
\lrag{z T} =
\lraug{\wp(0) F} = 
\int_0^\beta C_\gamma(0, t) \lraug{\frac{\part F}{\part\ow(t)}}dt
\ee{RecursionParts}
generates the relation \refp{zrecursion}.

Given $\ep \ge 0$, introduce the measure
\be
\dmuge = \frac{1}{\fz(\gep)} e^{-\ep^2 \int_0^\beta |\wp(s)|^2ds} \dmug\;,
\ee{masschange}
where the partition function is 
\be
\fz(\gep) = \lraug{e^{-\ep^2 \int_0^\beta |\wp(s)|^2ds}}
= \int_{\spt} e^{-\ep^2 \int_0^\beta |\wp(s)|^2ds} \dmug\;.
\ee{masspartitionfunction}
Also define the expectation
\be
\lrauge{F} = \int_\spt F \dmuge\;.
\ee{bracketlrauge}

\mbni{Proposition III.5. (Mass Renormalization)} 
{\it Let $\gamma=e^{-m\beta + i\w\th}$, where $|\gamma|<1$, and 
let $\ep>0$. Also let 
$m'=\sqrt{m^2+\ep^2}$ and $\gamma'= e^{-m'\beta+i\w\th}$. 
Then }
\be
\dmuge = \dmugp\;.
\ee{renormalization}
{\it Furthermore,}
\be
\fz(\gep) 
= \frac{|1-\gamma\phantom{'}|^2 }{|1-\gamma'|^2 } e^{\beta(m-m')}\;.
\ee{perturbedz}

\mbni{Proof.} The measures $\dmuge$ and $\dmugp$ are both Gaussian 
probability measures with mean zero. Therefore the measures agree if 
and only if they have the same covariance matrix. We use the integration 
by parts identity \refp{ibps} to prove this fact.  Start from this 
identity applied to $\lrauge{\overline{\wp(t)}\wp(s)}$, yielding
\be
\lrauge{\overline{\wp(t)}\wp(s)}
= \cmug - \ep^2 \int_0^\beta  C_\gamma(t,u) 
\lrauge{\overline{\wp(u)}\wp(s)} du\;.
\ee{covparts}
This can also be written  
\be
(I+\ep^2 C_\gamma) \lrauge{\overline{\wp(\cdot)} \wp(s)}
= C_\gamma(\cdot, s)\;,
\ee{firstidentity}
with $C_\gamma= (-D_\th^2+m^2)^{-1}$. The solution to this linear equation is 
the covariance $C_{\gamma,\ep }$ of $d\mu_{\gep}$ with the kernel
$C_{\gamma,\ep}(t,s)=\int_\spt \overline{\wp(t)}\wp(s) \dmuge 
=\lrauge{\overline{\wp(t)}\wp(s) }$.
Then 
\be
C_{\gamma,\ep} 
= (I + \ep^2 C_\gamma)^{-1} C_\gamma
= (C_\gamma^{-1} + \ep^2)^{-1}
=(-D_\th^2 +m^2+\ep^2)^{-1}
=C_{\gamma'}\;,
\ee{secidentity}
as claimed.

As a consequence of the mass renormalization formula, and with $H$ given in 
\refp{ham}, we have the identity
\be
\fz(\gep)
=\frac{\Tr_\fh\l(U(\th)^* e^{-\beta(H +\ep^2|z|^2)}\r)}
{\Tr_\fh\l(U(\th)^* e^{-\beta H}\r)}
\;.
\ee{renormalizedtrace}
Furthermore by Proposition II.1,
\be
\Tr_\fh\l(U(\th)^* e^{-\beta(-\part\opart + {m'}^2|z|^2 -m')}\r) 
= \frac{1}{|1-\gamma'|^2}\;,
\ee{perturbedtrace}
so \refp{perturbedz} follows.

\subsection{Gaussian Normal Ordering of Operator Products}

Each Gaussian functional $\lraug{\ \cdot\ }$ on coordinates $z(t)$ and 
$\oz(s)$ defines a ``normal-ordering'' of polynomials in the coordinates. 
In the case that this functional is the expectation on the zero-particle
state in $\CH$, namely for the case that $q=0$, the normal-ordered
product coincides with the Wick-ordered product. We denote the 
normal-ordering of $\oz(t)^r \, z(t)^s$ by $ \lrdq{ \oz(t)^r\, z(t)^s } $.
Define 
\be
c_\gamma = C_\gamma(0)>0\;,
\ee{constant}
where the positivity is a consequence
of \refp{covequaltime}. Let  
\be
\lrdq{\oz(t)^r\, z(t)^s} 
= \sum_{j=0}^{\min\{r,s\}} 
(-1)^j {{r}\choose {j}}{{s}\choose {j}}
\; j!\, c_\gamma^j \,\oz(t)^{r-j} \, z(t)^{s-j}\;.
\ee{normalorder}
In particular, a few of these normal-ordered monomials are
\beq
\lrdq{z(t)^n} &=& z(t)^n\;, \qquad \lrdq{\oz(t)^n} = \oz(t)^n\;,  \\
\lrdq{|z(t)|^2} &=& |z(t)|^2 - c_\gamma\;, \label{square}\\ 
\lrdq{|z(t)|^4} &=&|z(t)|^4 -4c_\gamma|z(t)|^2 +2c_\gamma^2\;,\label{fourth}\\
\lrdq{|z(t)|^6} &=& |z(t)|^6 -9c_\gamma|z(t)|^4 +18c_\gamma^2|z(t)|^2
     -6c_\gamma^3\;.\label{sixth}
\eeq
Note that ${{r}\choose {j}}^2 j! = ({{r}\choose {j}}j!)^2\frac{1}{j!}$,
so another expression for the normal-ordered monomial is 
\be
\lrdq{\oz^r\, z^s}  = e^{-c_\gamma \part\opart} \oz^r\, z^s\;.
\ee{exponentialordering}
This latter definition preserves its form when extended linearly 
to a general class of functions of $z$ and $\oz$ at equal time.
For $\gamma\neq 0$, the constant $c_\gamma$ has a 
limit as $m\to0$ with $\th$ fixed (see below); on the other
hand, the no-particle expectation of $|z|^2$ diverges. For this reason, we
conclude that the qualitative behavior of normal-ordered monomials depends 
on whether $\gamma=0$. We are interested in certain aspects of the 
infra-red ($\gamma\to0$) limit.

A principle property of these normal-ordered monomials is their 
orthogonality in the sense that 
\be
\lraug{ \lrdq{ \oz(t)^r z(t)^s } \; \lrdq{ \oz(t')^{r'} z(t')^{s'} } }
= \delta_{r s'}\delta_{s r'} \, r! \, s! \, 
C_\gamma(t-t')^{r} C_\gamma(t'-t)^{s}\;,
\ee{orthogonalmonomials}
and in particular 
\be
\lraug{ \lrdq{ \oz(t)^r z(t)^s } } =0\;.
\ee{notrace}
Also for $r=s$ and $r'=s'$,
\be
\lraug{ \lrdq{ |z(t)|^{2r}} \;  \lrdq{ |z(t')|^{2r'}} }
= \delta_{r r'} r!^2 |C_\gamma(t-t')|^{2r}\;.
\ee{orthogonalsize}

\subsection{Gaussian Normal Ordering on Path Space}
There is an equivalent notion of Gaussian normal-ordering of functions of 
paths $\wp(t)$ on $\spt$. In this case, normal ordering of a monomial with
respect to the measure $\dmug$ arises as the orthogonalization of the 
monomial to polynomials of lower degree in $\wp$ or $\ow$. This leads to 
polynomials in $\wp$ and $\ow$.  In particular, with the definition
\be
\Delta_{\wp} = 
\int_{0\le t,s \le\beta} C_\gamma(t-s) \frac{\part }{\part \wp(t)}
\frac{\part }{\part \ow(s)} dt ds\;,
\ee{functionallaplace}
we can define 
\be
\lrdq{F(\wp, \ow) } = e^{-\Delta_{\wp}} F(\wp,\ow) \;.
\ee{normalpath}
Then some normal-ordered (Hermite) polynomials are
In particular, a few of these normal-ordered monomials are
\beq
\lrdq{\wp(t)^n} &=& \wp(t)^n\;, \qquad \lrdq{\ow(t)^n} = \ow(t)^n\;,  \\
\lrdq{|\wp(t)|^2} &=& |\wp(t)|^2 - c_\gamma\;,\\
\lrdq{|\wp(t)|^4} &=& |\wp(t)|^4 -4c_\gamma|\wp(t)|^2 +2c_\gamma^2\;,\\
\lrdq{|\wp(t)|^6} &=& |\wp(t)|^6 -9c_\gamma|\wp(t)|^4 +18c_\gamma^2|\wp(t)|^2
     -6c_\gamma^3\;,
\eeq
etc. The integrals of these polynomials satisfy
\be
\lraug{ \lrdq{ \ow(t)^r\, \wp(t)^s }\; \lrdq{ \ow(t')^{r'}\, \wp(t')^{s'} } }
= \delta_{r s'}\delta_{s r'} \, r! \, s! \, 
C_\gamma(t-t')^{r} C_\gamma(t'-t)^{s}\;.
\ee{worthogonalmonomials}

\subsection{The Zero-Mass Limit ($m \to 0$)}
We consider here the $m\to 0$ limit with the inverse temperature $\beta$ 
fixed. In terms of $\gamma$ this entails the radial limit $|\gamma|\to 1$.  
Neither the expectation \refp{qfunctional}, nor the coordinates \refp{coord} 
are defined in this limit.  However, the measure $d\mu_\gamma$ is 
well-defined.  The following result, combined with the identity of 
Proposition III.3, allows us to extend the functional \refp{qfunctional} 
to the case $m=0$.  Recall the definition of $\sing$ in \refp{sing},
as well as the normalizing factor $\fz_\gamma$ given in \refp{qfunctional}.

\mbni{Proposition III.6.} {\it Let $\gamma = e^{-m\beta+i\w\th}$, and 
$|\gamma|<1$. 
\begin{itemize}
\ritem{a.} There is a constant $M$, independent of $m, \beta,\th$ 
such that the partition function $\fz_\gamma$ satisfies 
\be
0 < \fz_\gamma \le \frac{M}{\beta^2m^2}\;.
\ee{UniformZBound}
\ritem{b.} If $\th\not\in\sing$, then the measures $d\mu_\gamma(\wp(\cdot))$ 
converge weakly as measures on $\spt$ as $m\to0$ with $\beta, \th$ 
fixed. 
\ritem{c.}  The expectations 
$\lra{ \lrp{ \oz(t_1)\oz(t_2)\cdots 
\oz(t_n) z(s_1)z(s_2)\cdots z(s_m)}_+}_\gamma$ 
converge as $m\to0$ with $\beta, \th$ fixed. 
\end{itemize}
}

\mbni{Remark 1.} For a particular phase $e^{i\w\th}$  
the nuclear space $\st $ is independent of $m$.  Thus we can formulate 
continuity and convergence of the measures $d\mu_\gamma(\wp(\cdot))$ 
on a fixed space $\spt$. If we allow the phase $e^{i\w\th}$ 
of $\gamma$ to vary,
then we must formulate a more general notion of continuity and convergence
involving a family of measures on a family of spaces.  In this broader 
context convergence of moments, satisfying some uniform bound, provides a 
natural framework. We have investigated such convergence in a 
related problem some time ago \cite{convergence}, but we do not discuss these 
questions further here.

\mbni{Remark 2.} Let $m=0$, and let $\gamma=e^{i\w\th}\neq 1$. By 
comparing Proposition II.3, Proposition II.5 and \refp{pair}, 
we evaluate the kernel of $C_{\gamma}$ as 
\be
C_{\gamma}(t,s) = \frac{\beta}{4\sin^2(\frac{\w\th}{2})} 
%= \beta  \sum_{n = -\infty}^\infty \frac{1}{(2\pi n+ \w\th)^2}\;.
\ee{radial}
In fact, \refp{radial} is independent of $t$ and $s$,
so
\be
\lra{f,C_{\gamma} g}_{L^2} 
= \frac{\beta^3}{4\sin^2(\frac{\w\th}{2})} \overline{\hat f(0)} \hat g(0)\;.
\ee{limitingcov}

\mbni{Proof.} The bound \refp{UniformZBound} follows from the explicit
form of $\fz_\gamma$ established in Proposition II.1, as well as the
bound $|1-\gamma| > {\rm const.}\, \beta m$.  The bound of part (a) follows.

For fixed $\w\th$ the space $\spt$ remains fixed. 
Gelfand and Vilenkin \cite{Gelfand} show in \S IV.4.2 that weak convergence
of Gaussian probability measures on a fixed nuclear space 
is equivalent to convergence of the first
and second moments as operators on the nuclear space.  In the $m\to0$ limit 
that we consider here, the eigenfunctions of $C_\gamma$ remain 
fixed, while the eigenvalues converge as $m\to0$,
\be
\beta \hat C(E) = {1\over E^2 +m^2} \to {1\over E^2} \;.
\ee{limitingspectrum}
For $\th \in (0,{2\pi\over \w})$ and bounded away from the
endpoints of the interval, $E$ is bounded away from zero.  Hence the 
operator $C_\gamma$ is norm convergent as $m\to 0$.

\subsection{The Zero-Temperature Limit ($\gamma\to 0 $)}
The massive zero-temperature limit is $\beta\to\infty$ with $m>0$ fixed, 
as $\beta$ denotes the inverse temperature.  In terms of $\gamma$, 
this is the limit $\gamma \to 0$. The limiting covariance is just the vacuum 
expectation value 
\be
\lim_{\gamma\to 0} C_\gamma(\xi) =  C_0(\xi)=\frac{1}{2m} e^{-m|\xi|}\;,
\ee{vevlimit}
and it does not depend on the twist angle $\th$. Interestingly, the vacuum 
expectation covariance \refp{vevlimit} has no zero-mass limit, and this is 
one aspect of the ``infra-red'' problem alluded to in \S I.  

There is a different 
normalization of the coordinate that corresponds to the standard 
normalization used in the discussion of the zero-momentum mode of 
a quantized string. Let 
\be
z_{\rm bare}=m^{1/2}z = \frac{1}{\sqrt2} (a_{+}^* +a_{-})
\ee{bare} 
denoting the coordinate similar to \refp{coord} but without the scaling
by $m^{-1/2}$.   
Then the twisted expectation of the time-ordered product of coordinates
$z_{\rm bare}$ and their conjugates do have a zero-mass limit, determined by
\be
\lim_{\gamma\to0} C_\gamma^{\rm bare}(\xi) 
= \lim_{\gamma\to0} \lra{ (\overline{ z}_{\rm bare}(t) z_{\rm bare}(s) )_+}
= C_0^{\rm bare}(\xi)=\frac{1}{2} e^{-m|\xi|}\;,
\ee{betamass}

Finally we remark that the $m\to0$ and the $\gamma\to0$ limits cannot be 
interchanged.  The $m\to0$ limit of the 
pair correlation function \refp{radial}
has no $\beta\to\infty$ limit.  Likewise, the $\gamma\to0$ 
(or $\beta\to\infty$) limit \refp{vevlimit} has no $m\to0 $ limit.

\section{The $n$-Component Oscillator}
\setcounter{equation}{0}
The oscillator treated above has a straight-forward generalization to 
an $n$-component oscillator with the coordinate taking values in $\C^n$. 
Likewise, there is an probability measure on vector valued paths, whose
integrals give rise to twisted Gibbs expectations with respect to the
$n$-component oscillator.  We a brief outline of these properties here.

\subsection{Complex Operator Coordinates}  
Let $z=\{z_1, z_2, \ldots, z_n\}\in\C^n$ denote this coordinate. We may 
express $z$ in terms of $2n$ independent annihilation operators $a_{j\pm}$, 
$j=1,\ldots,n$, and their adjoints,
\be
z_j = \frac{1}{\sqrt{2m}}(a_{j+}^* + a_{j-}) \;,\quad\hensp{and }\quad
\partial_j =  \sqrt{\frac{m}{2}}(a_{j+} - a_{j-}^*) \;.
\ee{ncomponentz}
Introduce the mutually commuting 
number operators
$N_+ = \sumj a_{j+}^*a_{j+}$ and $N_- = \sumj a_{j-}^*a_{j-}$. We express 
the Hamiltonian $H = H_0$ and the twist generator $J$ as 
\be
H_0 = \sumj m (N_{j+} + N_{j-}) \;, \quad\hensp{and }\quad
J = \sumj \wj (N_{j+} - N_{j-}) \;.
\ee{multicomponentHandJ}
We only consider equal masses here, but we allow for different twist 
weights $\wj$ for the twist in each coordinate direction. Then  
$\uth=e^{iJ\th}$ generates the transformation
\be
z_j \to \uth z_j \uth^* = e^{i\wj\th} z_j\;, 
\quad\hensp{that we also write as}\quad
\uth z \uth^* = e^{i\w\th} z\;.
\ee{jtwist}
Likewise
\be
\partial_j \to \uth \partial_j \uth^* = e^{-i\wj\th} \partial_j\;.
\ee{ptwist}
The corresponding expectations $\lra{\,\cdot\,}_\gamma$ on $L^2(\C^n, d^nz)$ 
depend on an $n$-component parameter $\gamma=\{\gamma_1, \ldots, \gamma_n\}$, 
with $\gamma_j=e^{-m\beta+i\wj\th}$. 
On account of Proposition III.6.a, the partition function satisfies
\be
0 < \fz_\gamma \le \frac{M^n}{m^{2n}}\;,
\ee{NPartitionFunction}
where $M$ is a constant independent of $m,n,\beta,\th$.  The 
expectation of one coordinate vanishes, and the expectation of
two coordinates equals
\be
C_\gamma(t,s) = \lra{\oz(t) z(s)}_\gamma\;,
\ee{ncomponentexpectations}
with $C_\gamma(t,s) = C_\gamma(\xi)$ a diagonal matrix, depending on the 
difference variable $\xi=t-s$.  The entries of this matrix are
$\{C_\gamma(\xi)_{ij}\}$, where $1\le i,j\le n$, and they equal 
\be
C_\gamma(\xi)_{ij}= \lra{\oz_i(t) z_j(s)}_\gamma
=\delta_{ij} C_{\gamma_j}(\xi)\;.
\ee{ccomponents}
Furthermore,
\be
C_\gamma(\xi + \beta)_{ij} 
  = \delta_{ij} e^{ -i\wj\theta} C_{\gamma_j}(\xi)\;.                         
\ee{nComponentPeriodicity}
Define the constant $c_j$ as the $j^{th}$-component of the covariance 
on the diagonal, namely
\be
c_j = C_{\gamma_j}(0)\;,
\ee{cj}
and let
\be
\fc = \sumj c_j\;.
\ee{lpconstant1}
More generally, let
\be
\fc^{(p)} = \sumj c_j^p\;.
\ee{lpconstant}

\subsection{Normal Ordering}
We revisit the combinatorics of normal ordering in the case of the 
multi-component oscillator.  Let the number of components be fixed at $n$. 
Introduce the Laplacian on $L^2(\C^n)$ as 
\be
\Delta_c = \sumj c_j \part_j \opart_j\;,
\ee{cnlaplace}
Then define normal ordering of a function $P(z,\oz)$ on 
$L^2(\C^n)$ by the Laplacian $\Delta_\gamma$ acting on $P$ as
\be                                                  
\lrdq{ P(z,\oz) } = e^{-\Delta_c}  P(z, \oz)\;.
\ee{ncomponentnormalorder}
As some particular examples, the following polynomials 
in $|z|^2=\sumj |z_j|^2$ are normal ordered:
\beq               
\lrdq{ 1 }  &=& 1\;,\\
\lrdq{ |z|^2 }  &=& |z|^2 - \fc\;, \label{z2}\\
\lrdq{ |z|^4 }  &=& (|z|^2)^2 - 2\fc |z|^2 + \fc^2 
              -2 \sumj c_j |z_j|^2 + \fc^{(2)} \;.\label{z4}
\eeq
If all the $\gamma_j$ are equal, then $c_j=c$ and these expressions simplify 
to 
\be
\lrdq{ |z|^2 }  = |z|^2 -n c\;,\\
\ee{equal2}
and
\be
\lrdq{ |z|^4 }  = |z|^4 - 2(n+1) c |z|^2 + n(n+1)c^2\;.
\ee{equal4}
These normal-ordered monomials have the orthogonality property
\be
\lra{ \lrdq{ |z(t)|^{2k} } \;  \lrdq{ |z(s)|^{2k'} } }_\gamma 
= \delta_{k,k'}  (k!)^2 \sum_{k_1+k_2+\cdots k_n =k} 
       \prodj |C_{\gamma_j}(t-s)|^{2k_j}\;,
\ee{nordered}
where $k_j\in \Z_+$ ranges over non-negative integers.
In particular \refp{equal2} and \refp{equal4} reduce in the case 
$n=1$ and $k=1,2$ to \refp{square}--\refp{fourth}, and \refp{nordered}
reduces in the case of equal $\gamma_j$'s to \refp{orthogonalsize}.

The path space for the $n$-component oscillator 
consists of vector-valued paths $\wp(t) = \{ \wgj \}$, that are 
parameterized by the set of numbers $\gamma_j = e^{-m\beta + i\wj\th }$.  
Thus ${\wp}: \R \to \cs'_{\th} \oplus  \cs'_{\th} 
\oplus \cdots \oplus \cs'_{\th}$, with the implicit notation that 
the $j^{\rm th}$ component $\wgj(t)$ of the path satisfies the periodicity 
condition 
\be
\wgj(t+\beta) = e^{i\wj \th} \wgj(t) \;.
\ee{multiperiod}
The measure on 
$\cs'_{\th} \oplus  \cs'_{\th} \oplus \cdots
\oplus \cs'_{\th}$ is the product measure 
\be
d\mu_\gamma(\wg(\cdot)) = \prod_{j=1}^n d\mu_{\gamma_j}(\wgj(\cdot))\;,
\ee{productmeasure}
and integrals with respect to this measure are related to the expectations 
in the twisted Gibbs functional $\lraug {  \, \cdot\,  } $ by a Feynman-Kac
formula similar to \refp{BasicFK} and \refp{fkidentity}.

\section{Non-Gaussian Functionals and Non-Gaussian Measures}
\setcounter{equation}{0}
The twisted Gibbs expectations in \S II--IV are Gaussian. They arise from
harmonic oscillators, and are they are characterized by a 
linear equation of motion.  In this section we construct non-Gaussian
Feynman-Kac probability measures providing Feynman-Kac representations for 
various twisted, non-Gaussian functionals.  These examples arise 
from non-linear, twist-positive quantum-mechanical systems 
$\{H,\uth,\fh\}$.   The twist-invariant Hamiltonians 
$H = H_0 + V$ are perturbations of the Hamiltonian $H_0$, the 
Hamiltonian of a massive, $n$-component harmonic oscillator. We take the 
perturbation $V=V(z,\oz)$ to be a twist-invariant, multiplication operator 
on $\C^n$. In the subsections that follow, we detail our assumptions on $V$,  
establish twist positivity, and establish the Feynman-Kac representation.  
For simplicity, we study polynomial potentials.  We also investigate 
some aspects of the $m\to0$ limit.

\subsection{Allowed Potentials}
\mbni{Definition V.1} a. {\it The potential function $V(z,\oz)$ is
{\rm allowed} if it is a real polynomial in $z$ and $\oz$ that is
bounded from below and twist-invariant, namely 
\be
\uth V(z,\oz) \uth^* = V(e^{i\wj\th}z_j, e^{-i\wj\th}\oz_j) = V(z,\oz)\;.
\ee{TwistInvariantPotential}
}

b. {\it The potential $V$ is {\rm elliptic}, if there are positive constants
$M_1, M_2 <\infty$ such that 
\be
|z|^2 \le M_1 (V(z,\oz) + M_2)\;,
\ee{EllipticityBound}
where $|z|^2$ denotes $\sumj |\zj|^2$.}

c. {\it A potential $V$ is {\rm infrared regular} if there are 
positive constants $M_1, M_2 <\infty$ such that the Laplacian of $V$ 
satisfies
\be
\l|\sumj\frac{\part^2 V(z,\oz)}{\part z_j \part\oz_j}\r| 
\le M_1 (V(z,\oz)  + M_2)\;.
\ee{IRBound}
}

In the introduction, we mentioned some examples of acceptable 
potential functions.  Clearly $|z|^2$ is
twist invariant, so any polynomial function of $|z|^2$ that is bounded from 
below is acceptable.  Another example mentioned in the 
introduction arose from the absolute square 
of a holomorphic, quasihomogeneous polynomial $W$. A holomorphic polynomial 
$W(z)$ is quasihomogeneous with positive weights $\{\wj\}$ for the 
coordinates $\{ \zj\}$ if 
\be
W(z) = \sumj \wj \zj \frac{\part W(z)}{\part\zj}\;.
\ee{qhInfinitesimal}
The identity \refp{qhInfinitesimal} is the infinitesimal form of the 
relation, 
\be
W(\{e^{i\wj\th}\zj \}) = e^{i\th} W(\{\zj \})\;.
\ee{qhIntegrated}
In other words, quasihomogenity with weights $\{\wj\}$ means that under 
twisting by $\uth$ (defined with weights $\{\wj\}$), 
\be
\uth W \uth^\ast = e^{i\th} W\;.
\ee{qhSymmetry}
A homogeneous, holomorphic polynomial has equal weights $\wj$, 
equal to the inverse of the degree of the polynomial. 
The absolute square $|W(z)|^2$ of a holomorphic, quasihomogeneous 
polynomial $W(z)$ is twist-invariant.  
Furthermore, if $W(z)$ is holomorphic and quasihomogeneous, 
then the $k^{\rm th}$-component of its gradient $\part W/\part z_k$
is also quasihomogeneous with weights $\{ \w_j(1-\w_k)^{-1} \}$. In fact 
\be
\uth \frac{\part W(z)}{\part z_k} \uth^* 
     = e^{i(1-\w_k)\th} \frac{\part W(z)}{\part z_k}\;,
\ee{derivativeTwist}
yielding the claimed quasihomogeneity. Thus the absolute square of 
the gradient of a holomorphic, quasihomogeneous polynomial
\be
V(z, \oz) = \sumj|\frac{\part W(z)}{\part\zj}|^2
\ee{GradSuperPotential}
is twist invariant.

Let us give some examples of holomorphic, quasihomogeneous polynomials 
$W(z)$ and some potentials $V(z,\oz)$.
For instance, choose for $W(z)$ to be a sum of monomials in the individual 
coordinates, 
\be
W(z) = \sumj c_j \frac{\zj^{n_j}}{n_j}\;, \quad\hensp{where}\quad  
1\le n_j\in\Z\;.
\ee{qhPotential1}
In this case, $W$ is holomorphic and quasihomogeneous with weights
$\wj = 1/n_j$.  The squared gradient 
$V= \sumj |\frac{\part W(z)}{\part\zj}|^2$ has the form
\be
V(z,\oz) = \sumj |c_j \zj^{n_j-1}|^{2}\;.
\ee{exampleV1}

A second example is 
\be
W(z_1, z_2) = z_1^k + z_1 z_2^l\;,
\ee{qhPotential2}
that is quasihomogeneous with weights $\w_1 = 1/k $ and $\w_2 = (k-1)/kl$.  
In this case the gradient squared has the form
\be
V(z,\oz) = |k z_1^{k-1} + z_2^l |^2 + l^2 |z_1|^2 |z_2|^{2(l-1)}\;.
\ee{exampleV2}

\subsection{Hamiltonians and the Trotter Product Formula}
Let $\D\in\fh$ denote the domain of $C^\infty$ functions of $z$ and $\oz$.
We say that the Hamiltonian $H=H_0 + V$, with $V$ an allowed potential, is
an {\it allowed Hamiltonian}, and it has the form
\be
H = -\part \opart + V_1(z,\oz)\;, \quad\hensp{where}\quad 
V_1(z,\oz) = m^2 |z|^2 + V(z,\oz) -nm\;.
\ee{}
Since $V$ is bounded below, it follows that $V_1$ is elliptic.  Such 
Hamiltonians $H$ are known to be essentially self adjoint, and to have 
trace-class heat kernels $e^{-\beta H}$, for all $\beta>0$.  

The Trotter product representation is a form of the Lie formula 
for semigroups such as  $e^{-\beta H}$ with unbounded generators,
\be
e^{-\beta H} 
= \slim{N\to\infty} 
\l(e^{- \beta H_0/2N} e^{- \beta V/N} e^{- \beta H_0/2N}\r)^N \;.
\ee{TrotterConvergence}
For operators $H_0$ and $V$ that are self adjoint and bounded from below, 
and such that $H=H_0+V$ is 
essentially self adjoint, convergence of 
the Lie-Trotter product formula \refp{TrotterConvergence} is standard; 
see for example, Theorem A.5.1 of \cite{QP}. We require a variation.

\mbni{Proposition V.2}  {\it Let $H$ be an allowed Hamiltonian. Then 
the Trotter product representation holds in the form} 
\be
\lim_{N\to\infty}
\Trh{\uth^*  \l( e^{- \beta H_0/2N} e^{- \beta V/N} 
            e^{- \beta H_0/2N} \r)^N }
 = \Trh{\uth^* e^{-\beta H}} \;.
\ee{TraceConvergence}
\mbni{Remark.} The important point is that we
can interchange the order of the limit over $N$ and the trace as claimed in 
\refp{TraceConvergence}.  
Let us define
$T_N = e^{- \beta H_0/2N} e^{- \beta V/N} e^{- \beta H_0/2N}$. Clearly 
$T_N$ is self adjoint and positive. It is no loss of generality to add
a constant to $V$ so that $0\le V$, and we make this assumption in proving
the proposition. Therefore
\be
0\le T_N \le e^{-\beta H_0/N}\;,
\ee{TNBound} 
so $T_N$ is bounded from above by the contraction semigroup generated 
by $H_0$. Rewrite \refp{TrotterConvergence} as
\be
e^{-\beta H} =\slim{N\to\infty} T_N^N\;.
\ee{TrotterConvergence2}
and  \refp{TraceConvergence} as 
\be
\Trh{\uth^* e^{-\beta H}} = \lim_{N\to\infty} \Trh{\uth^*  T^N_N }\;.
\ee{TraceConvergence2}

\mbni{Proof.}  
Let us denote the 
$j^{\rm th}$-eigenvalue of $T_N$ by $e^{-\la_{N,j}/N}$, where we order
the eigenvalues so that $\la_{N,j}$ is an increasing function of $j$. 
Then the eigenvalues of $T_N^N$ are then $e^{-\la_{N,j}}$.
Likewise, denote the eigenvalues of $e^{-\beta H_0}$ by $e^{-\la_{j}}$,
with $\la_j$ increasing.  From the inequality \refp{TNBound} and the 
minimax principle, we infer that $\la_j \le \la_{N,j}$.  Furthermore,
from the explicit form of the eigenvalues of $H_0$ with mass $m$ and for 
$z\in \C^n$, there is a constant $M>0$ such that 
$\la_j \ge Mm j^{1/2n}$. Hence we conclude that there is a minimum
rate of decay for the $j^{\rm th}$-eigenvalue of $T_N^N$, namely
\be
e^{-\la_{N,j}} \le e^{-\la_j} \le e^{Mmj^{1/2n}}\;.
\ee{EigenvalueUniformDecay}
As a consequence, given $\ep>0$, there exists $J = J(\ep)<\infty$, 
independent of $N$, such that for every $N$, 
\be
\sum_{j=J}^\infty e^{-\la_{N,j}} < \frac13 \ep\;.
\ee{TailBound}

Observe that $\uth^* T_N = T_N\uth^* $, so $\uth^* $ can be diagonalized 
simultaneously with each $T_N^N$. Let $e^{i\delta_{N,j} -\la_{N,j}}$ 
denote the spectrum of $\uth T_N^N$ in the corresponding orthonormal
basis $|N,j\rangle$.  We compute the trace of $\uth^*  T^N_N$ in this basis,
and the trace of $\uth^*  e^{-\beta H}$ in its basis 
$|j\rangle$ corresponding to eigenvalues $e^{i\delta_j -\la_j}$.  Thus using
\refp{TailBound}, we have 
\beq
&&\hskip -.75in \l| \Trh{\uth^*  e^{-\beta H}} 
- \Trh{\uth^* \l(e^{- \beta H_0/2N} e^{-\beta V/N} e^{-\beta H_0/2N} \r)^N }
                \r|\nn\\
&& \hskip 1.1in = \l|\Trh{\uth^*  T_N^N} -\Trh{\uth^* e^{-\beta H}}\r| \nn\\
&& \hskip 1.1in \le \frac23 \ep
   + \l| \sum_{j=0}^J \l(e^{i\delta_{N,j} - \la_{N,j}} 
     -  e^{i\delta_{j} - \la_{j}} \r) \r| \;.
\label{FiniteDimensionalReduction}
\eeq
Since $T^N_N$ converges strongly to $e^{-\beta H}$ as $N\to\infty$, 
we can use standard contour integral methods to establish that each
eigenvector $|N,j\rangle$ converges strongly to the eigenvector
$|j\rangle$ for the limiting operator. It follows that each 
eigenvalue $e^{i\delta_{N,j} -\la_{N,j}}$ converges
to $e^{i\delta_{j} -\la_{j}}$. Hence the finite number of eigenvalues in the
sum on the right of \refp{FiniteDimensionalReduction} converge 
as $N\to\infty$, uniformly for $0\le j\le J$.  We therefore may choose 
$N_0(\ep)$ so that for $N>N_0(\ep)$,
\be
\l| \sum_{j=0}^J \l(e^{i\delta_{N,j} - \la_{N,j}} 
     -  e^{i\delta_{j} - \la_{j}} \r) \r| \le \frac13 \ep\;.
\ee{MainConvergence}
We conclude that we have shown that given $\ep>0$, there exists $N_0$ such
that for $N>N_0(\ep)$,
\be
\l|\Trh{\uth^*  e^{-\beta H}}
-  \Trh{\uth^* \l( e^{- \beta H_0/2N} e^{- \beta V/N} 
          e^{- \beta H_0/2N} \r)^N }  \r| < \ep\;.
\ee{EndOfConvergenceProof}
This completes the proof of the proposition.

Note that in the notation \refp{timeorder} 
of products that are time-ordered with respect
to the action of the semigroup $e^{-\beta H_0}$, we could also write
\be
\Trh{\uth^*  T_N^N}
= \Trh{\uth^*  \l(\prod_{j=1}^N 
e^{-\frac{\beta}{N}
V(z((j-\frac12)\beta/N),\oz((j-\frac12)\beta/N))}
    \r)_+ e^{-\beta H_0}}\;.
\ee{TimeOrderedFormOfTrotter}
Thus we can interpret the convergence \refp{} as a convergence of 
expectations of time-ordered products.  We need a more general form of
convergence of time-ordered products. 

\mbni{Proposition V.3} {\it Let $H$ be an allowed Hamiltonian.  Then }
\beq
&&\hskip -.37in
\Trh{\uth^*  \l(\oz(t_1)\cdots \oz(t_r)z(s_1)\cdots z(s_{r'})\r)_+ 
      e^{-\beta H}}\nn\\
&&\hskip .02in =\lim_{N\to\infty} 
   \Trh{\uth^*  \l(\oz(t_1)\cdots \oz(t_r)z(s_1)\cdots z(s_{r'})
      T_N^N \r)_+ }\nn\\
&&\hskip .02in =\lim_{N\to\infty} 
   \Trh{\uth^*  \l(\oz(t_1)\cdots \oz(t_r)z(s_1)\cdots z(s_{r'})
      \l( \prod_{j=1}^N e^{-\frac{\beta}{N}
      V(z((j-\frac12)\beta/N),\oz((j-\frac12)\beta/N))}
    \r) \r)_+ e^{-\beta H_0}} ,\nn\\
\label{TrotterExpectations}
\eeq
{\it where time ordering in the first term of \refp{TrotterExpectations}
is defined by $e^{-\beta H}$, while in the other two terms 
it is defined by $e^{-\beta H_0}$.}

The proof of this proposition is a variation on the proof of Proposition
V.2, so we omit the details.  

\subsection{Non-Gaussian Feynman-Kac Representations}
With $V$ equal to one of our allowed potentials, consider  
$H = H_0 + V$. In general the corresponding twisted Gibbs functional 
\refp{TwistedGibbsFunctional} is non-Gaussian.  Recall our definitions 
of the twisted partition function and the twisted relative partition 
function,
\be
\fz_\gamma = \Tr_\fh \l(\uth^*  e^{-\beta H_0}\r)\;, \quad\hensp{and}\quad
\fz^V_\gamma = \frac{\Tr_\fh \l(\uth^*  e^{-\beta H}\r) }
                {\Tr_\fh \l(\uth^*  e^{-\beta H_0}\r)  }\;.
\ee{gammaZ}
From Proposition II.1, generalized to the case $z\in\C^n$,
\be
\fz_\gamma = \prod_{j=1}^n |1-\gamma_j|^{-2}\;.
\ee{zGaussian}

\mbni{Proposition V.4.}  {\it Let $H = H_0 + V$. Then  

{\bf a. } The twisted trace of $e^{-\beta H}$ has the representation
\be
\Tr_\fh\l(\uth^*  e^{-\beta H}\r) 
= \fz^V_\gamma\, \fz_\gamma\;,
\ee{TwistedTrace}
where the twisted relative partition function has the representation}
\be
\fz^V_\gamma = \int e^{-\int_0^\beta V(\wp(s),\,\overline{\wp(s)})ds}
     d\mu_{\gamma} > 0\;.
\ee{relativeZ}

{\bf b.} {\it The twisted Gibbs functional $\lra{\ \cdot\ }_\gamma^V$ 
defined by 
\be
\lra{\ \cdot\ }_\gamma^V
= \frac{\Tr_\fh\l(U(\th)^* \ \cdot\ e^{-\beta H}\r)}
{\Tr_\fh\l(U(\th)^* e^{-\beta H}\r)}
\ee{TGFH}
has a Feynman-Kac representation given by the measure 
\be
d\mu_\gamma^{V} 
  = \frac{e^{-\int_0^\beta V(\wp(s), \,\overline{\wp(s)})ds}
          d\mu_\gamma}
     {\int e^{-\int_0^\beta V(\wp(s), \,\overline{\wp(s)})ds}
            d\mu_\gamma}\;.
\ee{ExampleFK}
Thus for the time-ordered product coordinates $T$ defined by 
\be
T= (\oz_{j_1}(t_1) \cdots \oz_{j_r}(t_r) 
    z_{j'_1}(s_1) \cdots z_{j'_{r'}}(s_{r'}) )_+\;,
\ee{TOPExpectation}
and for the corresponding function of paths $X$ defined by 
\be
X(\wp,\ow)= \overline{\wp(t_1) \wp(t_2) \cdots \wp(t_n)} 
\wp(s_1) \wp(s_2)\cdots \wp(s_r) \;,
\ee{TOPFunction} 
the expectations satisfy 
\be
\lra{\ T\ }_\gamma^V
= \int \,X(\wp,\ow) \, d\mu_\gamma^{V} \;.
\ee{NonGaussianFK}
}

\mbni{Proof.}  Use Propositions V.2 and V.3 to write 
the twisted Gibbs functional in the form 
\be
\frac{\Tr_\fh\l(U(\th)^* \ T\ e^{-\beta H}\r)}
{\Tr_\fh\l(U(\th)^* e^{-\beta H}\r)}
=\lim_{N \to\infty} 
 \frac{\Tr_\fh\l(U(\th)^* \ T\ \l(e^{- \beta H_0/2N} e^{- \beta V/N}
        e^{- \beta H_0/2N}\r)^N\r) }
 {\Tr_\fh\l(U(\th)^*  \l(e^{- \beta H_0/2N} e^{- \beta V/N}
      e^{- \beta H_0/2N} \r)^N\r)}\;.
\ee{TrotterGibbs}
We can now rewrite the expressions in the numerator and the denominator
using Proposition III.3 (in the $n$-dimensional case). In order to apply the
Gaussian Feynman-Kac formula, we not only need expectations
of time-ordered products of coordinates provided by the proposition as such,
but we also need the Gaussian representation of bounded exponential functions 
$e^{-V(z(t))}$ by the corresponding function of paths, $e^{ -V(\wp(t))}$.  
We obtain this extension from the general functional analysis of measues.

Let us consider first the denominator of \refp{TrotterGibbs}. Using 
Proposition III.3 we obtain, for the normalized twisted expectation 
\be
\frac{\Trh{ U(\th)^* \l(e^{- \beta H_0/2N} e^{- \beta V/N}
       e^{- \beta H_0/2N} \r)^N } }
   {\Trh { \uth^*  e^{-\beta H_0} }}
= \int e^{- \frac{\beta}{N} \sum_{j=1}^{N} 
         V(\wp((j+\frac12)\beta/N), 
           \,\overline{ \wp((j+\frac12)\beta/N) } ) } 
        d\mu_\gamma\;. 
\ee{DenominatorFK}
The factor ${\Tr_\fh \l(\uth e^{-\beta H_0}\r)}$ is the partition function
$\fz_\gamma$, and normalizes the expectation \refp{DenominatorFK}.  The same
factor arises in the numerator of \refp{TrotterGibbs}, and the two 
partition functions exactly cancel. Using Proposition V.2, and also 
using cyclicity of the trace to replace 
$\l( e^{- \beta H_0/2N} e^{- \beta V/N} e^{- \beta H_0/2N}\r)^N $ by 
$\l( e^{- \beta H_0/N} e^{- \beta V/N}\r)^N $, we have shown that
\beq
\Tr_\fh(U(\th)^* e^{-\beta H}) 
&=& \lim_{N\to\infty}
  \Tr_\fh \l(U(\th)^* \l(e^{- \beta H_0/N} e^{- \beta V/N}\r)^N \r) \nn\\
&=& \fz_\gamma\,\lim_{N\to\infty}
 \int e^{- \frac{\beta}{N}\sum_{j=1}^{N} 
         V(\wp(j\beta/N), \,\overline{\wp(j\beta/N)}) } 
        d\mu_\gamma\nn\\
&=& \fz_\gamma\, \int e^{-\int_0^\beta V(\wp(s),\,\overline{\wp(s)})ds}
     d\mu_{m,\beta,\theta} > 0\;.
\label{TrotterDenominator}
\eeq
In the last equality we use that fact that $V$ is bounded from below, so 
the exponential on the right side of \refp{TrotterDenominator} 
is bounded uniformly in $N$. It also converges pointwise as $N\to\infty$.  
Thus the integral converges by the dominated convergence theorem.  

Note that
twisted positivity for $\{H,\uth,\fh\}$ is a consequence of this 
Feynman-Kac representation, along with twist positivity for 
$\{H_0,\uth,\fh\}$. We have now proved part a. of the proposition. 
In order to establish part b. of the proposition, we apply
Proposition V.3 to expand the numerator of \refp{TrotterGibbs}.
By a similar argument to the above, we obtain 
\beq
\Trh{\uth^*  T e^{-\beta H}} 
&=& \fz_\gamma \lim_{N\to\infty}
 \int X(\wp,\ow)e^{- \frac{\beta}{N}\sum_{j=1}^{N} 
         V(\wp((j+\frac12)\beta/N), \,\overline{\wp((j+\frac12)\beta/N)}) } 
        d\mu_\gamma\nn\\
&=& \fz_\gamma\, \int X(\wp,\ow)
       e^{-\int_0^\beta V(\wp(s),\,\overline{\wp(s)})ds}
     d\mu_{m,\beta,\theta} \;.
\label{TrotterNumerator}
\eeq
Taking the ratio of \refp{TrotterNumerator} and 
\refp{TrotterDenominator}, we complete the proof of
Proposition V.4.

\subsection{The Zero-Mass Limit, Revisited}
It is here that we make distinctions between the general allowed
potentials in Definition V.1, and those that are also elliptic or 
infrared regular. Naturally, we obtain stronger results with 
more assumptions on $V$. Recall the definition of $\sing$ in \refp{sing}.

\mbni{Proposition V.5.}  Let $V$ be an allowed potential.  Consider the
limit as $m\to0$ with $\beta$ and $\th$ fixed. 
\begin{itemize}
\ritem {a.} If $\th\not\in\sing$, then $d\mu^V_\gamma$ converges
weakly.
\ritem {b.} If $\th\in\sing$, there exists a 
constant $M = M(\beta,V) <\infty$ such that 
\end{itemize}
\be
\Trh{\uth^*  e^{-\beta(H_0 + V)} }
 = \fz^V_\gamma   \fz_\gamma
  \le \frac{M}{m^{2n}}\;.
\ee{generalSingularity}

\mbni{Proof.}  A general potential $V$ is bounded from below, so 
$e^{-\int_0^\beta V ds}$ is bounded from above and is independent of $m$.  
In Proposition III.6 we showed that in the case $n=1$, 
the measure $d\mu_\gamma$ converges as $m\to0$ with fixed 
$\beta$ and $\th\not\in\sing$. For vector-valued paths that we consider
here, a similar argument shows that the product measure $d\mu_\gamma$ 
defined in \refp{productmeasure} converges.  
Thus $d\mu^V_\gamma$ converges by the dominated convergence theorem.
Furthermore for any $\th$, we have the elementary bound
\be
e^{-\int_0^\beta V(\wp(s),\ow(s))ds} d\mu_\gamma
\le \l( \sup_{z}\{e^{-\beta V(z,\oz)}\} \r)\, d\mu_{m,\beta,\th}\;,
\ee{generalElemBound}
so using Proposition II.1, we have 
\be
\fz^V_\gamma \le (\sup_{z}e^{-\beta V(z,\oz)})\,\fz_\gamma 
   = (\sup_{z}e^{-\beta V(z,\oz)})\, \prod_{j=1}^n|1-\gamma_j|^{-2}\;.
\ee{freePartitionBound}
As $m\to0$ with $\beta$ and $\theta$ fixed, 
$|1-\gamma_j| > {\rm const.}\,m\beta$.  The stated bound 
\refp{generalSingularity} then follows.

\mbni{Proposition V.6.}  {\it Let $V$ be an allowed, elliptic potential.
\begin{itemize}
\ritem {a.} Let $0<\la \le 1$ and $H(\la) = H_0 + \la^2V$.
Then there exists a constant $M = M(V) <\infty$, independent of 
$m, \la, \beta, \th$,  such that 
\be
\Trh{\uth^*  e^{-\beta H(\la)}}
 = \fz^{\la^2 V}_\gamma\, \fz_\gamma
  \le \l(\frac{M}{\beta\,(m+\la)}\r)^{2n}\;.
\ee{ellipticSingularity}
\ritem {b.} Suppose in addition that the polynomial $V$ is infrared regular. 
Consider the limit $m\to0$, with $\beta$,  $\th$, (and $\la=1$) fixed. 
Then the heat kernel $e^{-\beta H}$ converges in trace norm, and consequently
$\fz^V_\gamma \fz_\gamma$ converges as $m\to0$.  Furthermore 
the measure $d\mu^{V}_\gamma$ converges weakly as $m\to0$.
\end{itemize}
}
\mbni{Remark.} Let $||| \ \cdot \ |||_p$ denote the Schatten $p$-norm,
\be
||| T |||_p = \Tr_\fh\l( (T^*T)^{p/2}\r)^{1/p}\;,
\ee{Schatten}
with $p=1$ the trace norm, $p=2$ the Hilbert-Schmidt norm, 
and $p=\infty$ the operator norm.

\mbni{Proof.}  From Proposition V.1.a we infer the identity of 
$\Tr\l(\uth^*  e^{-\beta H(la)}\r) $ with 
$\fz^{\la^2 V}_\gamma\, \fz_\gamma$. 
Let us establish the bound \refp{ellipticSingularity}.  
For an elliptic potential, 
\be
\sumj |\wpj(s)|^2 \le M_1(V(\wp(s), \overline{\wp(s)})ds +M_2 )\;.
\ee{EllipticPathBound}
Thus writing $|\wp(s)|^2 = \sumj |\wpj(s)|^2 $, we infer 
\be
\fz^{\la^2 V}_\gamma\,  \fz_\gamma
  = \int_{\cs'} e^{-\int_0^\beta \la^2 V(\wp(s), \overline{\wp(s)})ds} 
        d\mu_\gamma \,\fz_\gamma
  \le \int_{\cs'} e^{\la^2 M_2\beta} e^{\la^2 M_1^{-1} 
    \int_0^\beta |\wp(s)|^2 ds} d\mu_\gamma \, \fz_\gamma\;.
\ee{MassStabilityBound}
Use the mass renormalization identity of Proposition III.5, with 
$\ep^2 = \la^2 M_1^{-1}$, and the identity \refp{zGaussian} to obtain 
as 
\beq
\fz^{\la^2 V}_\gamma\, \fz_\gamma &\le& \l(e^{\la^2 M_2\beta } 
    \fz(\gamma, \la M_1^{-1/2})\r)\, \fz_\gamma\nn\\
   &=& e^{\la^2 M_2\beta } e^{n\beta(m- m')} 
    \l(\prod_{j=1}^n \frac{|1-\gamma_j|^2}{|1-\gamma_j'|^2}\r)\,
                    \fz_\gamma\nn\\
   &=& e^{\la^2 M_2\beta } e^{n\beta(m- m')} 
    \l(\prod_{j=1}^n \frac{1}{|1-\gamma_j'|^2}\r)\;,
                                     \label{MassStabilityBound2} 
\eeq
where $m' = (m^2 + \la^2/ M_1)^{1/2}$ and 
$\gamma_j' = e^{-m'\beta + i\wj\th}$.  Then 
$|1-\gamma'| > {\rm const.} \,m'\beta > {\rm const.} \beta\,(m+\la)$.
Hence we have established the desired bound \refp{ellipticSingularity}.  

In order to deal with convergence as $m\to0$, we use the Feynman-Kac
representation to work with Hamiltonian estimates. We establish the  
{\it second-order estimate} that follows from our assumed infrared
regularity bound on the Laplacian of $V$.

\mbni{Lemma V.7} {\it Let $V$ be an allowed, elliptic, and infrared 
regular potential. Let $0\le M_3$ be sufficiently large that 
$1\le V + M_3$, and $M_3>M_2$, where $M_2, M_1$ are the constants in
\refp{IRBound}. Then with $M_4 = 2nm + M_1 + M_3 $,
and for any $f$ in the domain of $H$,
\be
||\Delta f||^2 + ||(V +M_3)f||^2 + m^4 ||\,|z|^{2}f||^2 
   \le 2||(H_0 + V + M_4)f||^2\;,
\ee{SOE}
where $\Delta=\sumj \part_j\opart_j$.}
\mbni{Remark.} We work on the domain $\D\times\D$ and expand. We use the 
{\it double commutator method}, to write cross terms as a sum of 
positive terms and lower order terms, see \cite{JaffeThesis} or 
\cite{holonomy}.  The identity
\be
D^*D X + XDD^* 
= D^* X D +  D X D^* + [D^*,[D,  X]]
\ee{DoubleCommutator}
gives the following double commutator identity for the Laplacian,
\be
-\Delta X -X\Delta 
= \sumj\l( \part_j^*X\part_j +  \part_j X\part_j^*\r)
    - \sumj [\opart_j,[\part_j, X]]\;.
\ee{LaplaceDoubleCommutator}

\mbni{Proof.}  Computing the square of $H + M_3 + nm$ we have
\beq
(H + M_3 + nm)^2 &=& (-\Delta +m^2|z|^2 +V +M_3)^2\nn\\
&=& (\Delta)^2 + (V+M_3 + m^2|z|^2)^2 \nn\\
&&\hskip .3in - \Delta(V+M_3 + m^2|z|^2) - (V+M_3 + m^2|z|^2)\Delta \nn\\
&=& (\Delta)^2 + (V+M_3)^2 + (m^2|z|^2)^2 + 2(V+M_3)(m^2|z|^2) \nn\\
&& \hskip.3in + \sumj (\part_j^* (V+M_3 + m^2|z|^2)\part_j  
+ \part_j (V+M_3 + m^2|z|^2) \part_j^*) \nn\\
&& \hskip .3in + \sumj \, [\part_j^* ,[\part_j , V+M_3 + m^2|z|^2]]\nn\\
&\ge& (\Delta)^2 + (V+M_3)^2 + m^4|z|^4 -nm^2
    - \sumj \, [\opart_j ,[\part_j ,V ]]\;.\nn\\
\label{SecondOrder1}
\eeq
The double commutator term is just the Laplacian of $V$, that obeys 
the infrared regularity bound \refp{IRBound}.  
\beq 
(H + M_3 + nm)^2 
&\ge& (\Delta)^2 + (V+M_3)^2 + m^4|z|^4 
    - nm^2 - M_1(V+M_2)\nn\\
&\ge& (\Delta)^2 + \frac12 (V+M_3)^2 + m^4|z|^4 
     -nm^2 -\frac12 M_1^{2}\nn\\
\label{SecondOrder2}
\eeq
Thus with $M_4 = 2nm + M_1 + M_3 $, we have 
\be
2(H + M_4)^2 
\ge (\Delta)^2 +  (V+M_3)^2 + m^4|z|^4 \;,
\ee{SecondOrder3}
Since $H$ is essentially self adjoint on $D$, the 
inequality \refp{SOE} follows.
\mbni{Proof of Proposition V.6.b}
Denote the $m$-dependence of $H$ by $\hm$, and add a constant to both
Hamiltonians so that they are positive.  Write
\beq
\ehm{\beta} - \ehmp{\beta} 
&=& \intbeta \ehm{(\beta-s)} (\hmp-\hm) \ehmp{s\beta} ds\nn\\
&=& \intbeta \ehm{(\beta-s)}\l( (m^{\prime}- m)(m^\prime +m)|z|^2 
      + n(m-m^\prime) \r)  \ehmp{s\beta} ds\;.
\label{Difference}
\eeq
Then using H\"older's inequality for Schatten norms,
\beq
&&\hskip -.3in |||\ehm{(\beta-s)} (\hmp-\hm) \ehmp{s\beta}|||_1\nn\\
&& \hskip .5in \le |||\ehm{(\beta-s)/2}|||_2 \,
         |||\ehm{(\beta-s)/2}(\hm + I)^{1/2}|||_\infty\nn\\
&&\hskip .8in \times |||(\hm + I)^{-1/2}
             \l( (m^{\prime}- m^)(m^\prime +m)|z|^2 
      + n(m-m^\prime) \r)(\hmp + I)^{-1/2}|||_\infty\nn\\
&&\hskip .8in \times |||(\hmp + I)^{1/2}\ehmp{s/2}|||_\infty \, 
        |||\ehmp{s/2}|||_2 \;.
\label{TraceBound1}
\eeq
We bound this using Lemma V.7 in the form 
$||(V+M_3)^{1/2}(\hm +I)^{-1/2}||\le 2^{1/4}$, along with the elliptic 
bound in the form $|z| (M_1 (V +M_2))^{-1/2}$, resulting in 
$|||z|(\hm +I)^{-1/2}||\le {\rm const.}$, and similarly with $m'$ replacing
$m$.  Thus assuming that $m,m'$ remain bounded, there is a constant 
independent of the parameters such that for $0<s<\beta$,
\be
|||\ehm{(\beta-s)} (\hmp-\hm) \ehmp{s\beta}|||_1
\le {\rm const.} (\beta -s )^{-1/2}\, s^{-1/2}\, |m-m'|\;.
\ee{TraceBound2}
Integrating this bound over $s$, we obtain the desired trace-norm convergence
of the heat kernel as $m\to 0$, namely the estimate
\be
|||\ehm{\beta} - \ehmp{\beta}|||_1 \le {\rm const.} |m-m'|\;.
\ee{TraceNormConvergence}

Using the Feynman-Kac representation of Proposition V.4.a,
\be
\Trh{\uth^*  e^{-\beta \hm}} = \fz^{(m)\,V}_\gamma\,\fz^{(m)}_\gamma\;,
\ee{MassDependentFK}
we infer the convergence the product $\fz^{(m)\,V}_\gamma\fz^{(m)}_\gamma$
as $m\to0$. This is the denominator in the representation \refp{TGFH} of the
twisted Gibbs functional, and it is the normalizing factor for the measure
\be
\fz^{(m)}_\gamma\, e^{-\intbeta V(\wp(s),\ow(s))ds} d\mu_\gamma\;.
\ee{unnormalizedmeasure}
After normalization , this measure is $d\mu^{(m)\, V}_\gamma$.
The proof of convergence of the integral of the measure 
\refp{unnormalizedmeasure} on products $X$ of coordinates \refp{TOPFunction}
proceeds similarly, and establishes the weak convergence of the 
perturbed measures $d\mu^{(m)\,V}$ as $m\to0$ as $m\to0$. We omit the details.

\section{Quantum Fields and Random Fields}
\setcounter{equation}{0}
In this section we generalize the construction in \S II--V 
to the case of an $n$-component, complex quantum field 
$\varphi(x) = \{\varphi_j(x)\}$. This time-zero field has a spatial 
coordinate $x$ lying in an $s$-torus $\torus$, so our space-time
is $\torus \times \R$.  Assume that the periods of the torus are
$\ell_i$, where $1\le i\le s$. The spatial volume is 
$\vol=\prod_{i=1}^s \ell_l$ Let us denote the lattice 
of momenta $k$ dual to $\torus$ as 
\be
\hattorus = \l\{k: k = \{k_1, k_2, \ldots, k_s\}\;, \ \hensp{where each}\  
  k_i = \frac{2\pi}{\ell_i}\Z\r\}\;,
\ee{DualLattice}
and let $kx=\sum_{i=1}^s k_i x_i$.

\subsection{Complex Free Fields}
The quantum mechanical Hilbert space on which the field acts 
is the on the Fock Hilbert space $\fh$ over the torus $\torus$.  
This Hilbert space is the tensor product of the Hilbert spaces for 
individual, non-interacting oscillators. Each oscillator has a frequency 
$\mu(k)=(k^2+m^2)^{1/2}$. Thus for $k\neq0$ the frequency does not vanish 
even in the limit that $m\to 0$.  Each component of the field has a Fourier
expansion 
\be
\varphi_j(x) = \frac{1}{\sqrt{\vol}}
 \l(z_j + \sum_{0\neq k\in\hattorus}\hat\varphi_j (k)e^{-ikx}\r)\;.
\ee{Fourierfield}
The constant Fourier modes $z_j$ are the coordinates considered
in \S II--V.  The non-constant Fourier modes have the form 
\be
\hat \varphi_j(k) 
= \frac{1}{\sqrt{2\mu(k)}}\l( a_{+,j}(k)^* + a_{-,j} (-k)\r)\;,
\ee{FourierComponents}
where we express them in terms of canonical annihilation and creation 
operators $a_\pm^\#$ that satisfy,
\be
[a_{\pm, j}(k), a_{\pm, j'}(k')^*] = \delta_{j,j'}\delta_{k,k'}I\;,
\ee{FockCCR}
and where all other commutators between pairs of $a_{\pm, j}^\#$'s vanish.

Define the mutually-commuting number operators $N_{\pm, j}(k)$ by 
\be
N_{\pm, j} (k) = a_{\pm, j}(k)^*a_{\pm, j}(k)\;.
\ee{NumberOperators}
Then express three fundamental operators in terms of the 
time zero Fourier components:
\beq
H_0 &=& \sumj \sum_{k\in\hattorus}
     \mu(k)\l(N_{+, j} (k) + N_{-, j} (k)\r)\;, \qquad
J = \sumj \sum_{k\in\hattorus} \wj \l(N_{+, j}(k) - N_{-,j} (k)\r)\;,\nn\\
&&\hensp{and}\ 
P_i = \sumj \sum_{k\in\hattorus} k_i\l(N_{+, j} (k) + N_{-, j} (k)\r)\;.
\label{FieldHamiltonian}
\eeq
These operators are the free-field Hamiltonian, the twist generator, and the 
components of the momentum operator respectively. 
With $x\in\torus$, and $xP = \sum_{i=1}^s x_i P_i$, let 
\be
\varphi(x+y) = e^{-ixP} \varphi(y) e^{ixP} \;,
\ee{spatialTranslation}

As we observed in the introduction, in the field case we introduce the
$(s+1)$-parameter symmetry group 
$\utth = e^{i\tau P +i\th J}$. This group acts on the components of the 
field as
\be
\utth \varphi_j(x) \utth^* = e^{i\wj\th} \varphi_j(x-\tau)\;,
\ee{FieldSymmetry}
and it implements a twist on Fourier components, 
\be
\utth \hat \varphi_j(k) \utth^* = e^{ik\tau  +i\wj\th} \hat \varphi_j(k)\;.
\ee{spatialFourierTranslation}
In keeping with the notation in \S II--IV, we define
$\overline{\varphi_j(x)} = \varphi_j(x)^\ast$, with the adjoint taken in the
sense of a densely-defined sesquilinear form on $\fh$.  Also for $t>0$, 
we define the imaginary time field $\varphi_j(x,t)$ as an operator with
domain equal to the range of $e^{-sH}$,  where $s>t$, namely 
\be
\varphi_j(x,t) = e^{-tH} \varphi_j(x) e^{tH}\;.
\ee{MoveField}
Let 
\be
\overline{\varphi_j}(x,t) = e^{-tH}\overline{\varphi_j}(x) e^{tH}
  = e^{-tH}\varphi_j(x)^\ast e^{tH}\;.
\ee{AdjointField}

In order to simplify our notation, we replace the parameter $\gamma$ used
in previous sections by a family of parameters $\gamma = \{\gamma_j(k) \}$ 
that contain information on the dependence of $\gamma$ on 
$j, k, \beta, \tau$, and  $\th$.  Set 
\be
\gamma = \{ \gamma_j(k)\}\;, \quad\hensp{where}\quad 
\gamma_j(k) = e^{-\mu(k)\beta + ik\tau  + i\wj\th}\;,
\ee{gok}
where $k\in\hat\torus$ and $1\le j\le n$.
Then we designate partition functions and expectations as before by 
$\fz_\gamma$ or $\lra{\ \cdot \ }_\gamma$, and thereby we designate
the dependence on all relevant variables.  
For example, we write the free-field twisted partition function as
\beq
\fz_{\gamma } 
&=& \Tr_\fh\l(\utth^* e^{-\beta H_0} \r)\nn\\
&=& \prod_{j=1}^n\prod_{k} \Tr_{\fh_{\{k,j,+\}}}
        \l( \og_j(k)^{N_{+, j} (k) }\r)
    \Tr_{\fh_{\{k,j,-\}}} \l( \gamma_j(-k)^{N_{-, j} (k) }\r)\;.
     \label{FieldPartitionFunction}
\eeq 
Here $\fh_{\{k,j,\pm\}}$ is the Hilbert space for 
the $\{k,j,\pm\}$-degrees
of freedom, and $\fh = \otimes_{\{k,j,\pm\}}\fh_{\{k,j,\pm\}}$.
In the case of paths we
retain the notation $\Phi_\tt$, and we also use 
$C_\tt$ for the pair correlation operator, 
in order to emphasize the dependence on these variables.

\mbni{Proposition VI.1} {\it With the above assumptions, the 
free field partition function is twist positive, 
\be
\fz_{\gamma } 
=\Tr_\fh\l(e^{-i\th J  -i\tau P -\beta H} \r)
= \prod_{j=1}^n \prod_{k\in\hattorus} \frac{1}{|1-\gamma_j(k)|^2} > 0\;.
\ee{FreeFieldTwistedPartition}
}

\mbni{Proof. } The trace factorizes, as indicated in 
\refp{FieldPartitionFunction}.  For a particular term in the 
product, perform the sum as in the proof of Proposition II.1. 
We obtain for a given $k$ and $j$, the contribution to the product equal to 
\be
(1-\og_j(k))^{-1} (1- \gamma_j(-k) )^{ -1}\;.
\ee{CalculateZMode}
Thus 
\be
\fz_{\gamma} 
= \prod_{k,j}(1 - \og_j(k))^{ -1} (1 - \gamma_j(-k) )^{ -1}\;.
\ee{CalculateZ}
This product converges as $|\gamma_j(k)| < e^{-\beta |k|}$.
Combining the result for modes $k$ and $-k$ yields the 
product \refp{FreeFieldTwistedPartition}, and the proof is complete.

\subsection{Twisted Expectations and The Pair Correlation Function}
Introduce the twisted expectation
\be
\lrag{ \ \cdot \  } 
  = \frac{\Tr_\fh\l(\ \cdot \  e^{-i\th J  -i\tau P -\beta H} \r)}
    {\Tr_\fh\l(e^{-i\th J  -i\tau P -\beta H} \r)}\;.
\ee{TwistedFieldExpectation}
Note that the expectation of one field vanishes, 
\be
\lrag{\varphi_i(x,t)}
= \lrag{\overline{\varphi_i}(x,t)} = 0\;.
\ee{SingleFieldExpectation}  
Furthermore,
\be
\lrag { \varphi_i(x,t) \varphi_j(y,s)} 
= \lrag { \overline{\varphi_i}(x,t) \overline{\varphi_j}(y,s)} = 0\;.
\ee{GaugeInvariance}
Define the pair correlation function 
$C_\tt(x-y, t-s)$ as the expectation of the time-ordered product of 
a field and a conjugate field.  In particular, let 
\be
C_\tt(x-y , t-s)_{ij} 
 = \lrag{(\overline{\varphi_i}(x,t)\varphi_j(y,s))_+} \;. 
\ee{PCFunction}
This function vanishes unless $i=j$. Furthermore the fields
$\varphi_j(x)$ are periodic with $x\in\torus$, so the pair correlation 
function has a partial Fourier representation 
\be
C_\tt(x-y , t-s)_{ij} 
 = \frac{1}{\vol}  \sumk \hat C_\tt(k;t-s)_{ij}e^{ ik(x-y)}\;,
\ee{PCFourierExpansion}
where $k$ ranges over the lattice \refp{DualLattice}.

\mbni{Proposition VI.2.} {\it The twisted Gibbs functional 
\refp{TwistedFieldExpectation} for the complex, massive, free scalar 
field on the space-time $\torus\times [0,\beta]$ is a Gaussian functional. 
The pair correlation has the Fourier representation 
\refp{PCFourierExpansion}, where 
\be
\hat C_\tt(k;t-s)_{ij} = \frac{1}{2\mu(k)}
\l(\l(\frac{\gamma_j(k)}{1-\gamma_j(k)} \r) e^{-\mu(k)(t-s)}
   + \l(\frac{\overline{\gamma_j(k)}}
           {1-\overline{\gamma_j(k)}} \r)e^{\mu(k)(t-s)}
   + e^{-\mu(k)|t-s|} \r)\delta_{ij}\;.
\ee{PCFourierCoefficients}
}

\mbni{Proof.} In order establish \refp{PCFourierCoefficients}, 
carry out the argument for each degree of freedom in exactly
the same fashion as for the analysis in \S II for the oscillator.  
This result generalizes \refp{kernel}. 
In the mode labelled by $k$ we replace $m$ by $\mu(k)$. Secondly,
we need to perform the twist arising from the $e^{ -i\tau P}$ group.
But for each $k$, this just modifies the angle $\wj\th$, replacing it
by the angle $\wj\th + k\tau $. With our definition \refp{gok} of 
$\gamma$ to include the proper parameter dependence, we obtain the Fourier 
coefficients \refp{PCFourierCoefficients} in a straightforward way.  
The effect of complex conjugating $\gamma_j(k)$ in one term gives the 
correct relation between the sign of the terms $\wj\th$ and  $k\tau $.
Once we have obtained t he formula for the pair correlation function, 
the proof of the Gaussian character of the functional \refp{fieldGibbs}
follows the proof of Proposition II.3.  We omit the details.

\mbni{Proposition VI.3} {\rm a.} {\it The pair correlation function and 
its Fourier coefficients satisfy twist relations. If 
$0\le t\le s\le \beta$ then
\be
C_\tt(x-y, t-s+\beta)_{jj} = e^{ -i\wj\th} C_\tt(x-y - \tau, t-s)_{jj}\;,
\ee{PCTwistRelation1}
while if $0\le s \le t \le \beta$, then 
\be
C_\tt(x-y, t-s -\beta)_{jj} = e^{i\wj\th} C_\tt(x-y + \tau, t-s)_{jj}\;.
\ee{PCTwistRelation2}

\qquad {\rm b.}  The kernel $C_\tt(x-y, t-s)_{ij}$ is hermitian in the sense
that 
\be
C_\tt(x-y,t-s)_{ij} = \overline{C_\tt(y-x, s-t)_{ji}} \;.
\ee{PCisHermitian}

\qquad {\rm c.} For $0\le t\le s\le \beta$, the coefficients in the Fourier 
representation of the pair correlation function satisfy 
\be
\hat C_\tt(k; t-s+\beta)_{jj} 
= e^{-i\wj\th -ik\tau } C_\tt(k; t-s)_{jj}\;.
\ee{PCTwistRelationF1}
On the other hand, for $0\le s\le t\le \beta$,
\be
\hat C_\tt(k; t-s-\beta)_{jj} 
= e^{-i\wj\th -ik\tau } C_\tt(k; t-s)_{jj}\;.
\ee{PCTwistRelationF2}

\qquad {\rm d.} As a consequence, each $\hat C_\tt(k;t-s)_{jj} $ 
has the representation
\be
\hat C_\tt (k; t-s)_{jj} 
= \sum_{E\in K_{\tt,j}} \frac{1}{E^2 +k^2 +m^2} 
            e^{iE(t-s)}\;,
\ee{PCSimpleForm}
where
\be
K_{\tt,j} =\frac1\beta \{2\pi\Z + \wj\th + k\tau  \}\;.
\ee{AllowedEnergies}
}
\mbni{Remark.} The original definition of the pair correlation function 
$C_\tt(x-y,t-s)$ is restricted to the domain $x-y\in\torus $ 
and $t-s\in [-\beta,\beta]$. The twist relation \refp{RFTwistRelation} 
provides a natural extension of $C_\tt(x-y,t-s)$ to $\torus \times \R$,
with \refp{PCTwistRelation1}, \refp{PCTwistRelation2}, and 
\refp{PCisHermitian} holding throughout.

\mbni{Proof.} The boundary condition for the pair correlation function
can be established by using the above definitions and cyclicity of the
trace. Consider the case $0\le t \le s\le \beta$. Then
\beq
C_\tt(x-y,t-s+\beta)_{jj}
&=& \lra{(\overline{ \varphi_j}(x,\beta-s+t)
     \varphi_j(y,0))_+}_{\gamma }\nn\\
&=& \lra{\varphi_j(y,0) 
      \overline{ \varphi_j}(x,\beta-s+t)}_{\gamma }\nn\\
&=& e^{ -i\wj\th} \lra{\overline{\varphi_j}(x,\beta-s+t)
           \varphi_j(y + \tau,\beta) }_{\gamma }\nn\\
&=& e^{ -i\wj\th} C_\tt(x - y - \tau, t-s)_{jj}\;. 
\label{PCTwistRelationDerived1}
\eeq
For the case $0\le s\le t\le\beta$, write 
\beq
C_\tt(x-y,t-s -\beta)_{jj}
&=& \lra{(\overline{ \varphi_j}(x,t-s)
     \varphi_j(y,\beta))_+}_{\gamma }\nn\\
&=& \lra{\overline{ \varphi_j}(x,t-s)
     \varphi_j(y,\beta)}_{\gamma }\nn\\
&=& e^{ i\wj\th}\lra{\varphi_j(y - \tau ,0) 
      \overline{ \varphi_j}(x,t-s)}_{\gamma }\nn\\
&=& e^{ i\wj\th} \lra{(\overline{\varphi_j}(x,t-s)
           \varphi_j(y - \tau,0))_+ }_{\gamma }\nn\\
&=& e^{ i\wj\th} C_\tt(x - y + \tau, t-s)_{jj}\;. 
\label{PCTwistRelationDerived2}
\eeq
Thus we have established \refp{PCTwistRelation1} and 
\refp{PCTwistRelation2}.

The hermiticity condition \refp{PCisHermitian} follows directly
from inspecting the representation 
\refp{PCFourierExpansion}--\refp{PCFourierCoefficients}.
In fact, \refp{PCFourierCoefficients} shows that $C_\tt(x-y,t-s)_{ij}$ 
is diagonal in the matrix index $ij$, so we write the representation
\refp{PCFourierExpansion} as 
\beq
\overline{ C_{\tt}(y-x, s-t)_{jj} } 
&=& \frac{1}{\vol}  \sumk \overline{ \hat C_\tt(k;s-t)_{jj} } 
    e^{ ik(x-y)}\nn\\
&=& \frac{1}{\vol}  \sumk  \hat C_\tt(k;t-s)_{jj} e^{ ik(x-y)}
= C_\tt(x-y, t-s)_{jj}\;.
\label{RequireHermite}
\eeq

We translate this boundary condition into a condition on the 
Fourier coefficients $\hat C_\tt (k; t-s)_{jj} $ by taking the Fourier series 
in the variable $x-y$.  This gives 
\refp{PCTwistRelationF1} and \refp{PCTwistRelationF2}.  
Thus
\be
\hat C_\tt (k; t-s)_{jj} = \sum_{E\in K_{\tt,j}} f(k; E) e^{iE(t-s)}\;.
\ee{PCCoefficientFourier}

Finally, as in \S II.6, we calculate the coefficients in the Fourier 
representation similarly to the proof of Proposition II.4.  The spectrum of 
$C_\tt$ then 
follows as claimed, namely \refp{FullCovarianceSpectrum} is a union of the 
contributions from the different components of the covariance operator on 
the matrix diagonal, and the individual spectra resulting from 
\refp{JointSpectrum}.

\subsection{The Free Field Pair Correlation Operator}
The pair correlation operator $C_\tt$ is the integral operator
with the pair correlation function as its integral kernel.
\be
(C_\tt f)_i(x,t) 
 =\sumj \int_{\torus \times [0,\beta]} C_\tt(x-y,t-s)_{ij} f_j(y,s) dy ds\;.
\ee{PCOperator}
We identify the pair correlation operator as the resolvent of a twisted 
Laplace operator $\Delta_\tt$.  Introduce the Hilbert space 
$\K = L^2(\CC) \otimes \C^n$ on which the pair correlation operator acts.
Here $\CC$ denotes the compactified space-time $\torus\times [0,\beta]$.
Let $\CS^{(j)}_\tt(\CC)$ denote functions on $\CC$ that have Fourier 
representations of the form 
\be
f(x,t) = \sum_{ {{E\in \frac{2\pi}{\beta}\Z} \atop {k\in\hattorus} }} 
\hat f(k,E)e^{ikx + iEt} e^{ -i(\wj\th +  k\tau )t/\beta}\;.
\ee{FourierFunctionsByj}
Here the set $\hattorus$ denotes the lattice 
$\prod_{j=1}^s\frac{2\pi}{\ell_j}\Z$ dual to the torus $\torus$. 
We assume that the coefficients $\hat f(k,E)$ decrease faster than the 
inverse of any polynomial function of $k^2 + E^2$.  The functions 
\refp{FourierFunctionsByj} satisfy the boundary condition 
\be
f(x,\beta) = e^{ -i\wj\th}f(x - \tau, 0)\;,
\ee{TwistedBoundaryConditionByj}         
relating the two ends of the cylinder.   
The space of $C^\infty$ functions $\CS^{(j)}_\tt(\CC)$ is dense in $L^2(\CC)$.
Endowed with the topology given by the countable set of norms 
\be
||f||_n = \sup_{k, E} (1 + |k|^2 + E^2)^n |\hat f(k,E)|\;, \hensp{for}
n\in\Z_+\;,
\ee{Norms}
the space $\CS^{(j)}_\tt(\CC)$ is a Schwartz space of $C^\infty$ functions.
Furthermore, the representation \refp{FourierFunctionsByj} shows that 
functions $f\in\CS^{(j)}_\tt(\CC)$ extend from 
$\CC = \torus\times [0,\beta]$ to $C^\infty $ function on $\torus\times\R$, 
that satisfy the {\it twist relation}
\be
f(x,t+\beta) = e^{i\wj\th}f(x+\tau, t)\;.
\ee{FunctionTwistRelation}

The functions $e^{ikx +iEt} e^{ -i(k\tau )t/\beta}e^{ -i\wj\th t/\beta} \in
\CS^{(j)}_\tt(\CC)$  are all simultaneous eigenfunctions 
of $-i\frac{\part}{\part t}$ and of $-i\frac{\part}{\part x_i}$. The 
joint spectrum of these two operators on these eigenvectors is the set 
\be 
K_{\tt,j} \times \hat\torus \;,
\ee{JointSpectrum}
where $K_{\tt,j} $ is defined in \refp{AllowedEnergies}.
As a consequence, the domain $\CS^{(j)}_\tt(\CC)$ is a domain of essential 
self-adjointness for the twisted Laplace operator
\be
\Delta^{(j)}_{\tt} = \frac{\partial^2}{\partial t^2} 
+ \sum_{1\le i\le s}\frac{\partial^2}{\partial x_i^2}\;.
\ee{Laplace}
The dense domain of definition 
\be
\CS_\tt = \CS^{(1)}_\tt(\CC)\oplus \CS^{(2)}_\tt(\CC) 
   \oplus\cdots\oplus\CS^{(n)}_\tt(\CC)
\subset \K
\ee{nCore}
provides a dense domain for the Laplace operator $\Delta_\tt$.
This is an $n\times n$ matrix with each entry an operator on $L^2(\CC)$. 
The entries are
\be
\{ (\Delta_{\tt})_{jj'} \} 
   = \{ \Delta^{(j)}_{\tt} \delta_{jj'} \}\;.
\ee{matrixLaplace}
Denote the corresponding resolvent by 
\be
(-\Delta_\tt +m^2)^{-1} \;.
\ee{spacetimeLaplace}
It has matrix elements 
\be
\{(-\Delta^{(j)}_\tt + m^2)^{-1}\delta_{jj'} \}\;,
\ee{spacetimeCovariance}
so that for $f_j\in L^2(\CC)$ and $f=\{f_j\}\in \K$,
\be
((-\Delta_\tt + m^2)^{-1} f)_j(x,t) 
    = \sum_{j'=1}^n \int_\CC (-\Delta_\tt + m^2)^{-1}(x-y, t-s)_{jj'} 
           f_{j'}(y,s) dy ds\;.
\ee{SpaceTimeCovarianceOperator}
We say for short that $(-\Delta_\tt + m^2)^{-1}(x-y, t-s)$ denotes 
the integral kernel for $(-\Delta_\tt + m^2)^{-1}$.
 
\mbni{Theorem VI.4.} {\it Consider the resolvent of the twisted Laplace 
operator $\Delta_\tt$ defined as the closure of the Laplace operator with 
the domain $\cs_\tt(\CC)$ and acting as \refp{SpaceTimeCovarianceOperator}.
This equals the pair correlation operator $C_\tt$ by the integral kernel
\refp{PCFourierCoefficients}. Namely}
\be
C_\tt = (-\Delta_\tt + m^2)^{-1}\;.
\ee{IdentityofTwists}
{\it Furthermore, the spectrum of $C_\tt$ is the set}
\be
(E^2 + k^2 +m^2)^{-1}\;, \quad\hensp{where}\quad 
E\in \cup_{1\le j\le n} K_{\tt,j} \;,\ \hensp{and}\  k\in\hat\torus\;,
\ee{FullCovarianceSpectrum}
{\it with $K_{\tt,j}$ defined in \refp{AllowedEnergies}.}

\mbni{Proof.}  The proof of the theorem follows the proof of Theorem
III.1.  The spectrum is given in Proposition VI.3.c.

Define the singular set $\sing = \sing(\Omega_j, \ell_j)$ appropriate for 
complex quantum fields by 
\be
\sing = \{ \tt: 0\in \cup_{1\le j\le n}\cup_{k\in\hat\torus} 
\{2\pi\Z - \wj\th - k\tau \} \}\;.
\ee{Fieldsing}
For $\tt\not\in\sing$, let
\be
M =\sup_{1\le j \le n} \sup_{E\in K_{\tt,j}, k\in\hat\torus} 
\frac{1}{E^2+k^2}
\ee{CBound}

\mbni{Corollary VI.5.} {\it Let $m>0$. The operator $C_\tt$ on $\K$ with 
integral kernel \refp{PCFourierCoefficients} is positive, and compact.
Assume also that $M<\infty$ and $\tt\not\in\sing$. Then the 
operator $C_\tt$ is norm convergent as $m\to 0$. }

\subsection{Random Fields}
In this section we discuss the appropriate path space for random
fields satisfying a twist relation \refp{RFTwistRelation} leading to a 
Feynman-Kac formula for quantum fields with the twist operator $\utth$. 
These random fields are just elements of $\CS'_\tt$, the dual space 
to $\CS_\tt$.  We let $\Phi_\tt(x,t)$ denote an element of $\CS'_\tt$.
The space $\CS'_\tt$ has a dense subspace of $C^\infty$
functions, 
\be
\CS_{\tau,-\th} \subset \CS'_\tt\;.
\ee{DualS}
We infer from Proposition VI.2 that the pair correlation operator 
$C_\tt$ maps an $\CS_\tt$ into itself, and in fact this map is continuous.  
Define the adjoint operator
\be
C^+_\tt : \CS'_\tt \rightarrow \CS'_\tt \qquad\hensp{by the relation}\qquad
(C^+_\tt \Phi_\tt)(f)  = \Phi_\tt(C_\tt f)\;.
\ee{DualCovariance}
It then follows that the kernel of $C^+_\tt$ satisfies the boundary condition
\be
C^+_\tt (x -y, t-s +\beta)_{jj} 
      = e^{ i\wj\th} C^+_\tt ( x - y - \tau, t-s)_{jj}\;.
\ee{T}
We write for the corresponding paths the relation \refp{RFTwistRelation}.

\subsection{The Gaussian Feynman-Kac Measure}
The twisted Gibbs state of a free field 
$\varphi(x) = \{\varphi_j(x)  \}$
has a Feynman-Kac representation with a Gaussian measure on $\CS'_\tt$. 
Define $d\mu_{\gamma }(\Phi_\tt)$ 
as the Gaussian, probability measure on 
$\CS'_\tt(\CC)$ with mean zero and covariance $C_\tt$.  Since we have
established that the twisted, free-field Gibbs functional is Gaussian,
we have as a consequence of the properties established in Proposition VI.4,

\mbni{Proposition VI.6.} a. {\it The twisted Gibbs functional for a free 
field $\varphi(x)$ with mass $m>0$, has a Feynman-Kac representation}
%\newpage
\beq
&& \lra{ \lrp{ \ov{j_1}{x_1}{t_1}\ov{j_2}{x_2}{t_2}\cdots\ov{j_r}{x_r}{t_r}
\v{j'_1}{x'_1}{t'_1}\v{j'_2}{x'_2}{t'_2}\cdots
\v{j'_{r'}}{x'_{r'}}{t'_{r'}}}_+}_{\gamma } \nn\\
&& \hskip .8in 
    =  \int_{\cs'_\tt(\CC)} \oP{j_1}{x_1}{t_1} \oP{j_2}{x_2}{t_2} \cdots
        \oP{j_r}{x_r}{t_r}\nn\\
&& \hskip 2.5in    
     \P{j'_1}{x'_1}{t'_1} \P{j'_2}{x'_2}{t'_2} \cdots
          \P{j'_{r'}}{x'_{r'}}{t'_{r'}} d\mu_{\gamma }(\Phi_\tt)\;.\nn\\
\label{FKFieldIdentity}
\eeq
\hskip .3in b. 
{\it Let $\tt\not\in\sing$ and let $M<\infty$. As $m\to0$, with 
$\beta,\tt$ be fixed, the measures $d\mu_{\gamma }(\Phi_\tt)$ 
converge weakly as measures on $\cs'_\tt(\CC)$, and the expectations 
\be
\lra{ \lrp{ \ov{j_1}{x_1}{t_1}\ov{j_2}{x_2}{t_2}\cdots\ov{j_r}{x_r}{t_r}
\v{j'_1}{x'_1}{t'_1}\v{j'_2}{x'_2}{t'_2}\cdots
\v{j'_{r'}}{x'_{r'}}{t'_{r'}}}_+}_{\gamma }
\ee{FieldExpectations}
converge as distributions in $\otimes^{r+r'}\CS'_\tt(\CC)$.}
%\hskip .3in c. 

\subsection{Non-Linear Quantum Fields and Non-Gaussian Measures}
Finally we remark on the construction of non-Gaussian, twist-invariant
Gibbs functionals for fields.  These measures give Feynman-Kac 
representations for twisted Gibbs functionals constructed from certain 
non-linear quantum fields.  In this paper we do not with to consider
ultraviolet questions, so we use a regularized interaction potential $V$. 
We follow the (standard) procedure introduced in \S V to construct 
Feynman-Kac representations for non-linear oscillators.  First we introduce
a regularized field $\varphi_{j,\chi}(x)$ from which we construct the 
non-linear interaction. Let
\be
\varphi_{j,\chi}(x) = \frac{1}{\sqrt{\vol}}
 \l(z_j + \sum_{k\neq 0}\hat\varphi_j (k) \chi(k) e^{-ikx}\r)\;, 
\ee{cutoffField}
where $0\le \chi(k) \le 1$, where $\chi(0) = 1$ and where $\chi(k)$ is 
rapidly decreasing as $|k|\to\infty$.  Likewise, let $\Phi_{\tt, \chi}(x,t)$ 
denote the random field $\Phi_\tt(x,t)$ after convolution 
in the spatial variable $x$ with the function whose
Fourier coefficients are $\chi(k)$.  Given a holomorphic, quasihomogeneous
potential function $W$ as in \S V, we form the perturbed Hamiltonian
\be
H = H_0 + \int_{\T^s} V(\{\varphi_{j,\chi}(x)\}, 
           \{\overline{ \varphi_{j,\chi}(x)}\})dx\;.
\ee{fieldPerturbationH}
This Hamiltonian then defines a twisted Gibbs expectation. It has a 
Feynman-Kac representation given by the measure
\be
d\mu_{\gamma }^{V}(\Ph(\cdot)) 
  = \frac{e^{-\int_\CC V(\Phi_{\tt,\chi}(y,s), 
    \overline{\Phi_{\tt,\chi}(y,s)}) ds dy }
d\mu_{\gamma }}
{\int e^{-\int_\CC V(\Phi_{\tt,\chi}(y,s), 
    \overline{\Phi_{\tt,\chi}(y,s)}) ds dy }
     d\mu_{\gamma }}\;.
\ee{FieldFKMeasure}
Since this measure is regularized, we can construct
it by methods similar to those in \S V.  We do not 
give the details, as we will return elsewhere to analyze 
such problems as well as to establish their dependence on the 
regularization.  As previously, we obtain

\mbni{Proposition VI.7.} a. {\it The $\utth$-twisted Gibbs functional for a 
field Hamiltonian $H=H_0+V$ has a Feynman-Kac representation given by the
measure \refp{FieldFKMeasure}.  
Let $\th\not\in\sing$ and let $\beta, \tt$ be fixed. As $m\to0$, 
The measures $d\mu^V_{\gamma }(\Phi_\tt)$ converge weakly in 
$\cs'_\tt(\CC)$, as do the expectations of products of fields. 
}

\subsection{Real Fields}
In order to treat real scalar fields $\varphi(x)$ it is sufficient to make
minor modifications of our preceding analysis. In particular, there is only
one set of creation and annihilation operators for each $\{k,j\}$. Thus in 
place of \refp{FourierComponents}  we have
\be
\hat \varphi_j(k) 
= \frac{1}{\sqrt{2\mu(k)}}\l( a_{j}(k)^* + a_{j} (-k)\r)\;.
\ee{RealFieldFourier}
As a consequence the translation group acts as a symmetry of $H_0$
as before, but there is no $\th$-twist symmetry.  We define the parameters
\be
\gamma = \{ \gamma(k)\}\;, \quad\hensp{where}\quad 
\gamma(k) = e^{-\mu(k)\beta + ik\tau }\;.
\ee{Realgok}
Then the modified results for the real case are 

\mbni{Proposition VI.8} a.  {\it The twisted partition function for 
the real, free scalar field is}
\be
\fz_\gamma = \Tr_\fh\l(e^{i\tau P -\beta H}\r)
= \prod_{k\in\hat\torus} 
       \frac{1}{|1-\gamma(k)|^{n}}\;.
\ee{RealZ}

b. {\it The real free-field pair correlation function is a multiple of the
identity in the matrix index,}
\be
C_\gamma(x-y,t-s)_{ij} 
=\lra{(\varphi_i(x,t) \varphi_j(y,s))_+}_\gamma 
= C_\gamma(x-y,t-s) \delta_{ij}\;,
\ee{PCFreeFieldDiagonal}
{and satisfies the twist relation}
\be
C_\gamma(x-y +\tau,t-s +\beta) = C_\gamma(x-y, t-s)\;.
\ee{PCRealTwist}

c. {\it The diagonal elements of the real, free-field pair correlation 
function have Fourier representation of the form 
\refp{PCFourierExpansion} with 
coefficients}
\be
\hat C_\tau(k;t-s) = \frac{1}{2\mu(k)}
\l(\l(\frac{\gamma(k)}{1-\gamma(k)} \r) e^{-\mu(k)(t-s)}
+ \l(\frac{\overline{\gamma(k)}}{1-\overline{\gamma(k)}} \r)
e^{\mu(k)(t-s)}  + e^{-\mu(k)|t-s|} \r)\;.
\ee{RealPCFourierCoefficients}

d. {\it The real, free-field pair correlation operator is the resolvent 
of the twisted Laplace operator $\Delta_\tau$ defined on the real
functions in $\CS_{\tau,0}(\CC)$,}
\be
C_\tau = (-\Delta_\tau + m^2)^{-1}\;.
\ee{RealCovariance}

e. {\it The real random paths $\Phi_\tau(x,t)$ satisfy}
\be
\Phi_\tau(x+\tau, t+\beta) = \Phi_\tau(x,t)\;.
\ee{RealTwistField}

As a consequence, a class of interacting Hamiltonians 
\be
H= H_0 + V
\ee{RealPerturbation}
exist, are twist positive, and have a Feynman-Kac representation with a
twist $U(\tau)$.  These interactions arise from real, 
translation-invariant, cutoff interactions with a potential $V$ that is 
bounded from below.  
The corresponding Feynman-Kac measure has the form,
\be
d\mu_{\gamma }^{V}(\Ph(\cdot)) 
  = \frac{e^{-\int_\CC V(\Phi_{\tau,\chi}(y,s)) ds dy }
d\mu_{\gamma }}
{\int e^{-\int_\CC V(\Phi_{\tau,\chi}(y,s)) ds dy }
     d\mu_{\gamma }}\;.
\ee{RealFKPerturbation}
We intend to return to these properties in another work in more detail.

\end{document}